\documentclass[%
 reprint,superscriptaddress,
 amsmath,amssymb,
 aps, physrev,
]{revtex4-2}

\usepackage{graphicx}
\usepackage[caption=false]{subfig}
\usepackage{lmodern}
\usepackage{dcolumn}
\usepackage{bm}
\usepackage[english]{babel}
\usepackage{siunitx}
\usepackage{hyperref}

\DeclareSIUnit\angstrom{\text {Å}}
\hypersetup{
  colorlinks   = true, 
  urlcolor     = white!30!black, 
  linkcolor    = white!30!black, 
  citecolor   = white!30!black 
}

\newcommand{\appref}[1]{Appendix~\ref{#1}}
\usepackage[commentmarkup=todo]{changes}
\usepackage[a4paper, left=2.5cm,right=2.5cm]{geometry}
\begin{document}

\preprint{APS/123-QED}

\title{\texorpdfstring{Probing the Fermi surface with Quantum Oscillation\\Measurements in the Dirac semimetal TaNiTe$_5$}{Probing the Fermi surface with Quantum Oscillation Measurements in the Dirac semimetal TaNiTe5}}

\author{Maximilian Daschner}
\email{maximilian.daschner@lmu.de}
\affiliation{Cavendish Laboratory, University of Cambridge, Cambridge CB3 OHE, United Kingdom}
\affiliation{Fakultät für Physik, Ludwig-Maximilians-Universität, München, Germany}

\author{Bruno Gudac}
\author{Mario Novak}
\affiliation{Department of Physics, Faculty of Science, University of Zagreb, Zagreb, Croatia}

\author{Cheng Liu}
\affiliation{Cavendish Laboratory, University of Cambridge, Cambridge CB3 OHE, United Kingdom}

\author{F. Malte Grosche}
\affiliation{Cavendish Laboratory, University of Cambridge, Cambridge CB3 OHE, United Kingdom}

\author{Ivan Kokanović}
\email{kivan@phy.hr}
\affiliation{Department of Physics, Faculty of Science, University of Zagreb, Zagreb, Croatia}
\affiliation{Cavendish Laboratory, University of Cambridge, Cambridge CB3 OHE, United Kingdom}

\begin{abstract}
\noindent
We report a detailed investigation of the Fermi surface in the layered Dirac semimetal TaNiTe$_5$. We probed the magnetization, magnetic torque and magnetoresistance in high-quality single crystals. Pronounced Shubnikov - de Haas (SdH) and de Haas - van Alphen (dHvA) oscillations are observed in magnetic fields above $\SI{3}{\tesla}$ and at temperatures of up to \SI{22}{\kelvin}. Multiple fundamental frequencies and light effective quasiparticle masses are obtained by fast Fourier transformation (FFT) and Lifshitz-Kosevich (LK) formula fits. The high resolution of the low-temperature FFT spectra allows us to investigate individual peaks in detail for the magnetic fields applied along all three crystallographic axes and the planes in between. Our investigation can confirm the density functional theory (DFT) calculated band structure and its corresponding Fermi surface.
\end{abstract}

\maketitle

\section{Introduction}
\noindent
Topological semimetals represent a new class of quantum materials hosting Dirac \cite{young2015dirac,yan17,xiong15,borisenko14} or Weyl fermions \cite{huang15,soluyanov15,weng15,lv15,xu15a,xu15b,xu15c,yang15,liu16}, in which the valence and conduction bands touch and form nodal points. Furthermore, Dirac nodal-lines in which Dirac cones form lines in reciprocal space lead to the more general concept of nodal-line semimetals (NLSMs) \cite{fang16,yu17,schoop16,fang15,bzdusek16,yang18,hu16}. NLSMs usually occur in three-dimensional materials where they hold more symmetries such as mirror and nonsymmorphic symmetries that protect the Dirac band touching structures against spin-orbit coupling (SOC) \cite{yang18,schoop16}. Evidence for robust Dirac nodal lines in (quasi) one-dimensional materials has been missing until recently when they were reported in the exfoliatable, in-plane anisotropic nonmagnetic semimetal TaNiTe$_5$ by combining angle-resolved photoemission spectroscopy (ARPES) and density functional theory (DFT) calculations \cite{hao2021multiple}. Quantum oscillation studies have reported evidence for Fermi surface pockets with nontrivial Berry phase in TaNiTe$_5$, and highly anisotropic properties were shown in magnetization and transport measurements \cite{xu2020anisotropic,chen21,ye22}. These previous studies were performed for magnetic fields applied parallel to the crystallographic \emph{a} and \emph{b} axes. However, a systematic study of quantum oscillations for magnetic fields parallel to the crystallographic \emph{c}-axis or in any of the planes perpendicular to the \emph{a} or \emph{b} axes remains absent. Here, we report a detailed investigation of the Fermi surface in TaNiTe$_5$ single crystals through magnetization, magnetic torque and magnetoresistance measurements for magnetic fields of up to \SI{14}{\tesla} along all three principal axes and a rotational study in all three planes.
\begin{figure*}[ht]
    \includegraphics[width=\textwidth]{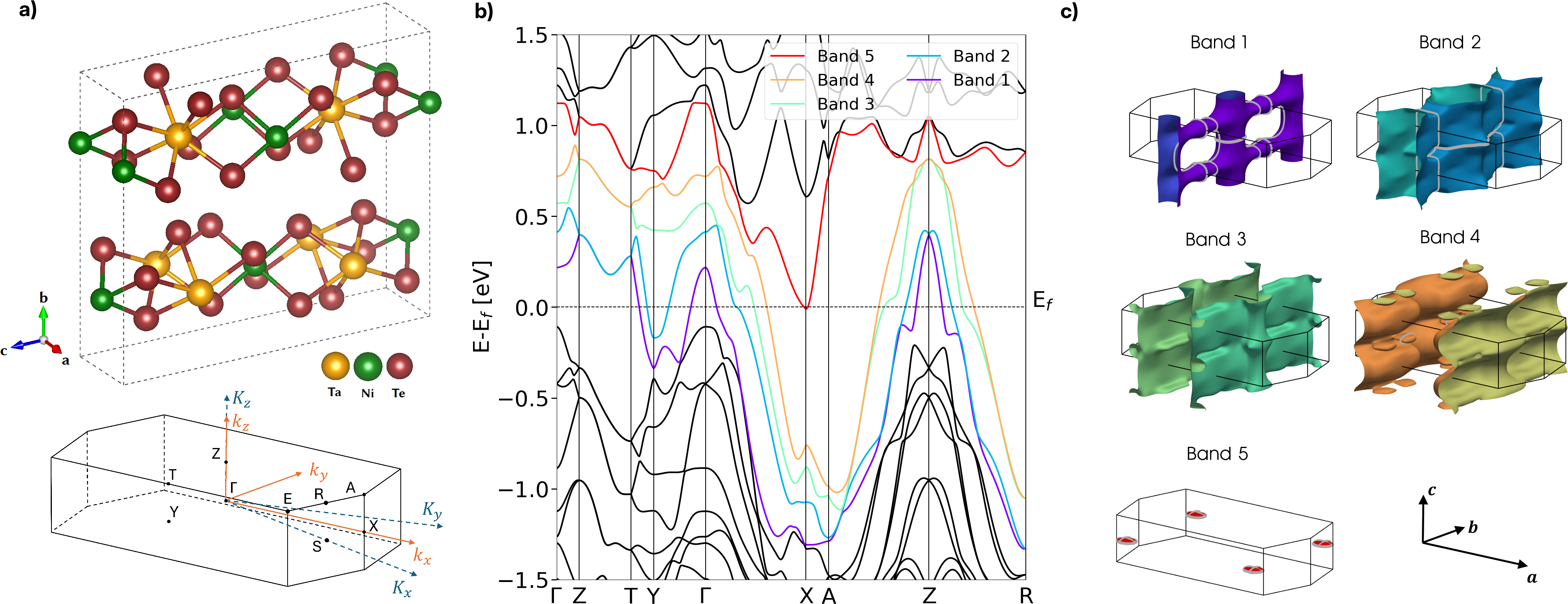}
    \caption{a) Unit cell and first Brillouin zone in TaNiTe$_5$. In the Brillouin zone, $k_x$, $k_y$, and $k_z$ point in conventional unit cell vector directions which go through high-symmetry points X, Y and Z and align with the crystallographic axes \textit{a}, \textit{b}, and \textit{c}, respectively, while $K_x$, $K_y$, and $K_z$ are the reciprocal lattice vectors that represent the periodicity of the Brillouin zone. b) Band structure including spin-orbit coupling along high-symmetry lines. c) Fermi sheets for TaNiTe$_5$ with the Fermi level shifted to lower energies by approximately \SI{20}{\milli\electronvolt}. Extremal orbits are shown for the magnetic field along $a$, $b$ and $c$ as gray lines in the first Brillouin zone.}
        \label{BZ_bandstructure}
\end{figure*}
We observed pronounced de Haas - van Alphen (dHvA) and Shubnikov - de Haas (SdH) oscillations with multiple frequencies at temperatures below \SI{22}{\kelvin} and magnetic fields above \SI{3}{\tesla}. The observed quantum oscillation frequencies in these crystals closely agree with each other, evidencing the same electronic structure. Changing the magnetic field direction results in a clear change of the oscillation frequencies, indicating anisotropic Fermi surface pockets. By analyzing the SdH and dHvA quantum oscillations, multiple fundamental frequencies, small effective masses, low Dingle temperatures, and high mobility of charge carriers are obtained. The high resolution of the low-temperature FFT spectra in the observed magnetization and magnetoresistance data allows us to investigate individual peaks in detail. Our results shed light on the electronic properties of the Dirac semimetal TaNiTe$_5$ and serve as a basis for the study of low-dimensional quantum materials and their topological properties.

\section{Experimental Methods}
\noindent
The TaNiTe$_5$ crystals were synthesized by the self-flux method \cite{chen21,liimatta89}. The mixture at a molar ratio of Ta:Ni:Te = 1:1:10 was loaded into an alumina crucible and sealed inside a quartz tube in high vacuum. The mixture was then heated to \SI{700}{\celsius} over the course of four days and slowly cooled to \SI{500}{\celsius} within a week. The resulting single crystals have an orthorhombic layered structure with the space group Cmcm (No. 63) \cite{liimatta89}. The structure features one-dimensional (1D) NiTe$_2$ chains along the crystallographic \emph{a}-axis, which form a quasi-2D layer by linking chains of Ta atoms along the \emph{c}-axis as can be seen in \autoref{BZ_bandstructure} a). The lattice constants are inferred from powder X-ray diffraction (XRD) data with $a = \SI{3.657}{\angstrom}$, $b = \SI{13.183}{\angstrom}$, and $c = \SI{15.119}{\angstrom}$. The peaks of the single-crystal XRD pattern showed no trace of an impurity phase, indicating high crystalline quality (see \appref{methods}). The crystals grow in the shape of flat needles along the \emph{a}-axis. Magnetization measurements in two crystals with masses of \SI{25}{\milli\gram} and \SI{27.4}{\milli\gram} were carried out using a 14 Tesla Quantum Design Physical Property Measurement System (PPMS) with VSM option. The magnetoresistance was measured using standard AC lock-in techniques in a 9 Tesla PPMS. We have also performed magnetic torque measurements on two crystals in magnetic fields of up to \SI{9}{\tesla} and \SI{13}{\tesla} for the angle-dependent measurement in the \emph{a}-\emph{b} plane and the \emph{b}-\emph{c} plane, respectively. These were carried out in the aforementioned 9 Tesla PPMS, and a customized 13 Tesla Oxford Instruments cryostat. The crystals were mounted on custom-made piezoresistive cantilevers and placed on a rotating platform.

\section{\emph{Ab initio} calculations}
\noindent

To gain insights into the electronic structure of TaNiTe$_5$, we have performed density functional theory (DFT) calculations using the full-potential linearized augmented plane wave method implemented in the Wien2k package \cite{blaha2001wien2k}. The PBE-GGA exchange-correlation functional \cite{Perdew96} was chosen here. The atomic sphere radii (muffin-tin radii) were chosen as $R_{mt} = 2.5$ a.u. for the Ta and Ni atoms, and $R_{mt}= 2.43$ a.u. for all Te atoms. The plane-wave cut-off parameter was chosen as $R_{mt}K_{max} = 8$, where $R_{mt}$ is the smallest atomic sphere radius in the unit cell and $K_{max}$ is the magnitude of the largest k-vector. Calculations were performed on a 59 × 59 × 13 k-point mesh in the full Brillouin zone for the band structure calculations shown here. \autoref{BZ_bandstructure} b) illustrates the resulting band structure including spin-orbit coupling (SOC) with its respective Fermi sheets shown in \autoref{BZ_bandstructure} c). The Fermi sheets for each band also include extremal areas for magnetic fields aligned with the three crystallographic axes \textit{a}, \textit{b} or \textit{c}.

From \autoref{BZ_bandstructure} b) it can be seen that linearly dispersing bands appear in pairs along the A-Z-R high-symmetry line, forming Type-I Dirac cones in Z, and Type-II Dirac cones in R as the bands maintain their spin-degeneracy. Similar to the material family members TaXTe$_5$ (X = Pd, Pt), TaNiTe$_5$ also hosts a four-fold degenerate nodal-line along Z-T, with linear dispersion along $\Gamma$-Z and T-Y. Symmetry considerations \cite{young2012dirac,young2015dirac} performed in the literature for materials that crystallise in the symmetry group Cmcm (63), e.g. TaNiTe$_5$ \cite{hao2021multiple}, TaPtTe$_5$ \cite{xiao2022dirac}, LaNiGa$_2$ \cite{badger2022dirac}, and Th$_2$BC$_2$ \cite{wang2022symmetry} show protection of such nodal-lines. Their energetic location is around \SI{200}{\milli\electronvolt} above the Fermi level, making it unlikely for them to influence our obtained experimental results \cite{xu2020anisotropic, jiao20,jiao21}.

\section{Theory}
In metals and semimetals, oscillations in various physical quantities in magnetic field are well described by the grand thermodynamical potential given by the Lifshitz-Kosevich (LK) formula \cite{shoenberg84}:

\begin{equation}\label{grand_thermodynamic_potential_LK}
\resizebox{0.88\linewidth}{!}{$\Omega = \sum\limits_{i} C_i B^{\frac{5}{2}} R_T R_D \cos\left(2\pi \left(\frac{F_i}{B}-\frac{1}{2}+\phi \right)\right)$}
\end{equation}

where the sum is over all orbits $i$ along a Fermi sheet, and we assume only one harmonic for simplicity. $B$ is the magnetic field, $C_i$ is a $B$-independent constant, $F$ is the dHvA frequency of the oscillations, which can be related to the extremal area $A$ of the individual pocket via the Onsager relation $F = A (\phi_0/2\pi^2)$, where $\phi_0$ is the magnetic flux quantum \cite{shoenberg84}, and $m^*$ is the quasiparticle effective mass. We also include a phase shift $\phi=\phi_B/2\pi+\delta$ where $\phi_B$ is the Berry phase, and $\delta$ is an additional phase shift depending on the dimensionality of the Fermi surfaces, i.e. $\delta$ = 0 for 2D Fermi surfaces, and $\delta$ = $\pm1/8$ for 3D Fermi surfaces (-1/8 for the electron-like and 1/8 for the hole-like). $R_T=X/\sinh(X)$ with $X=2 \pi^2 k_B m^* T/(\hbar eB)$ is the temperature damping factor, while $R_D=\exp(-X T_D/T)$ is the Dingle damping factor which accounts for finite particle lifetime.

The mean free path $l$ can be approximated under the assumption of a spherical Fermi surface as
\begin{equation}
	l = \frac{\sqrt{2e\hbar^3F}}{2\pi m^* k_B T_D}
\end{equation}
while quantum relaxation time and quantum mobility are defined as $\tau_q$ = h/(4$\pi^2$$k_B$T$_D$) and $\mu_q$ = e$\tau_q/m^*$, respectively.

Shubnikov - de Haas oscillations in the magnetoresistance are proportional to \autoref{grand_thermodynamic_potential_LK}, while de Haas - van Alphen oscillations in the magnetization can be derived via $M_{\parallel} = -\left(\frac{d\Omega}{dB}\right)$. Oscillations in the magnetic torque are then given by
\begin{equation}\label{magnetic_torque_definition}
	\mathbf{\tau} = \mathbf{M}_{\perp}\times \mathbf{B}
\end{equation}
where the perpendicular magnetization is given by
\begin{equation}\label{perp_magnetization}
	\mathbf{M_{\perp}} = -\frac{1}{F}\frac{dF}{d\theta}\cdot \mathbf{M_{\parallel}}
\end{equation}

Using this formalism, we will analyze the magnetization, magnetic torque and magnetoresistance data obtained for TaNiTe$_5$ in what follows.

\section{Experimental Results}
\subsection{Magnetization}
\noindent

\begin{figure}[t]
        \includegraphics[width=0.46\columnwidth]{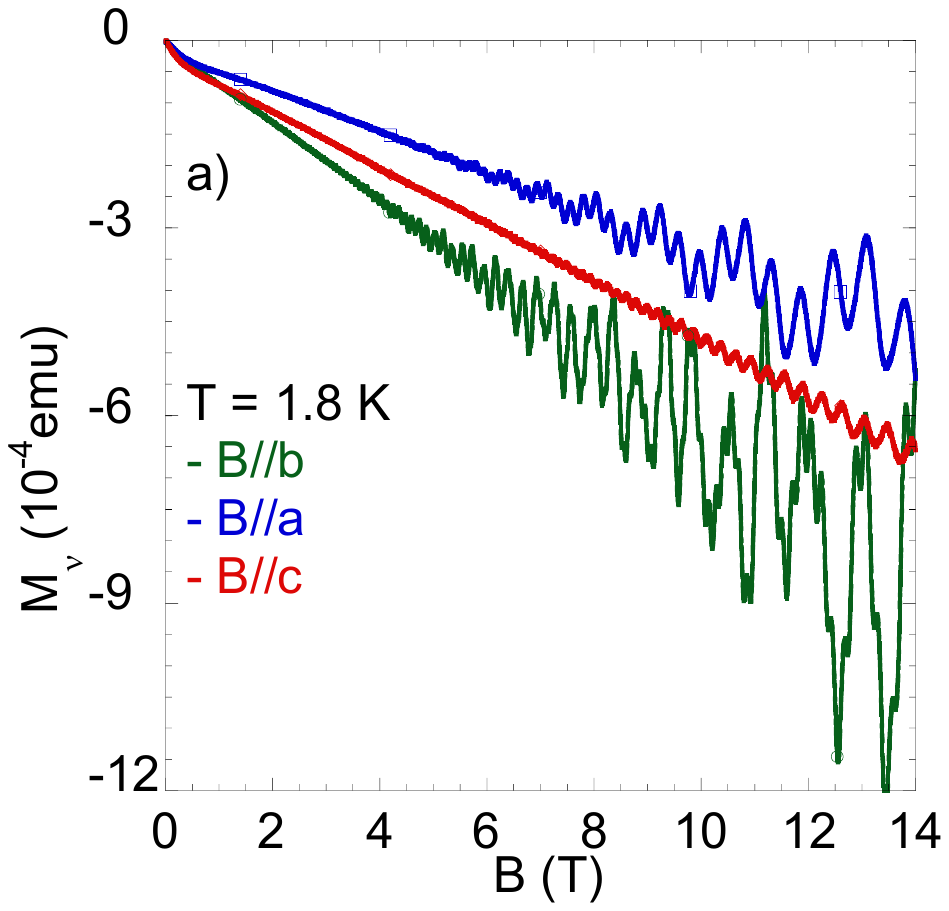}
        \includegraphics[width=0.52\columnwidth]{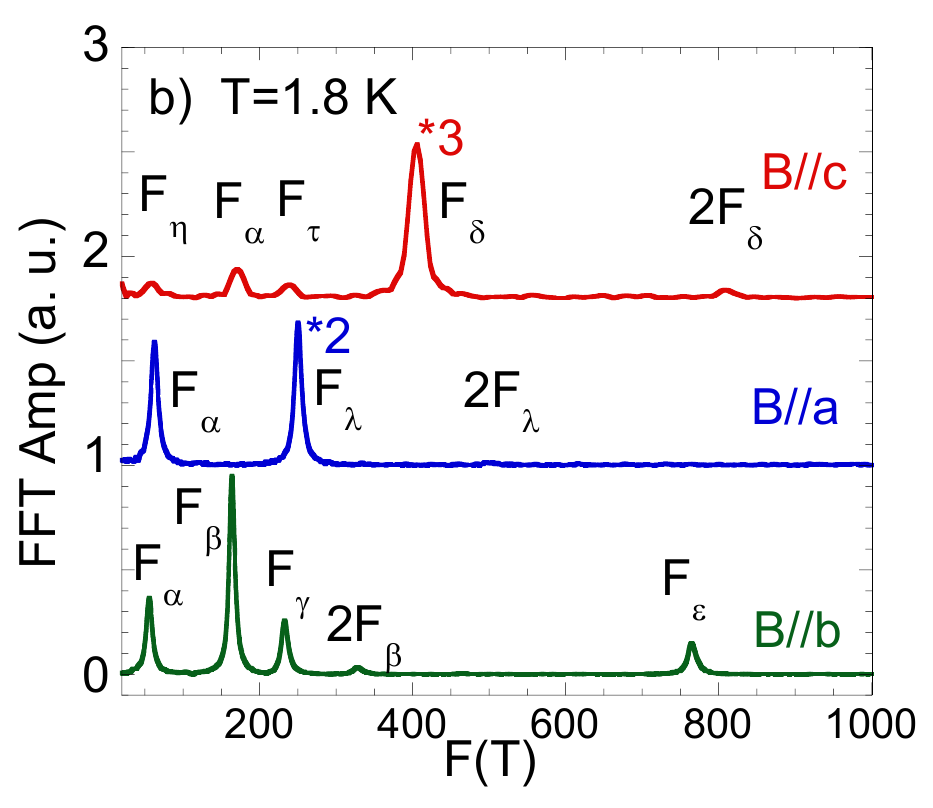}
	\caption{a) Magnetization of a layered TaNiTe$_5$ single crystal with a mass of 25.6 mg at a temperature of \SI{1.8}{\kelvin}. The magnetic field is applied parallel to the crystallographic \emph{b}, \emph{a}, and \emph{c} axes, respectively. The curves show an interpolation of data points with a few data points highlighted for \textbf{B}$\parallel$\emph{c} (red circle), \textbf{B}$\parallel$\emph{a} (blue square), and \textbf{B}$\parallel$\emph{b} (green triangle). b) FFT spectra of the dHvA oscillations in the isothermal magnetization data extracted after subtracting the diamagnetic backgrounds shown in a). The FFT amplitudes of the spectra have been generated from the magnetization data between \SI{4}{\tesla} and \SI{14}{\tesla} and are shifted vertically for clarity.}
    \label{magnetization_overview}
\end{figure}

\autoref{magnetization_overview} a) shows a series of magnetization measurements in TaNiTe$_5$ at a temperature of \SI{1.8}{\kelvin}, with magnetic fields applied along the crystallographic \emph{a}, \emph{b}, and \emph{c} axes. TaNiTe$_5$ exhibits large dHvA quantum oscillations with multiple frequencies superimposed on different linear diamagnetic backgrounds. By changing the direction of the magnetic field from \textbf{B}$\parallel$\emph{b} to \textbf{B}$\parallel$\emph{a}, and to \textbf{B}$\parallel$\emph{c}, the oscillation components are gradually suppressed, illustrating the anisotropic electronic properties of this material with an anisotropic magnetic susceptibility ratio of $\chi_a$:$\chi_c$:$\chi_b$=$3.3:4.5:6.3$ at \SI{1.8}{\kelvin}.

In \autoref{magnetization_overview} b), we have performed a FFT analysis of the isothermal magnetization data shown in \autoref{magnetization_overview} a). Several clearly resolved peaks are visible for the magnetic field parallel to each crystallographic axis, which we label with greek letters. The observed frequencies ($F$) in the FFT spectra are related to the extremal area through the Onsager relation. Frequencies we found here are consistent with the literature \cite{xu2020anisotropic,chen21,ye22}.

\begin{figure}[ht]
            \includegraphics[width=\columnwidth]{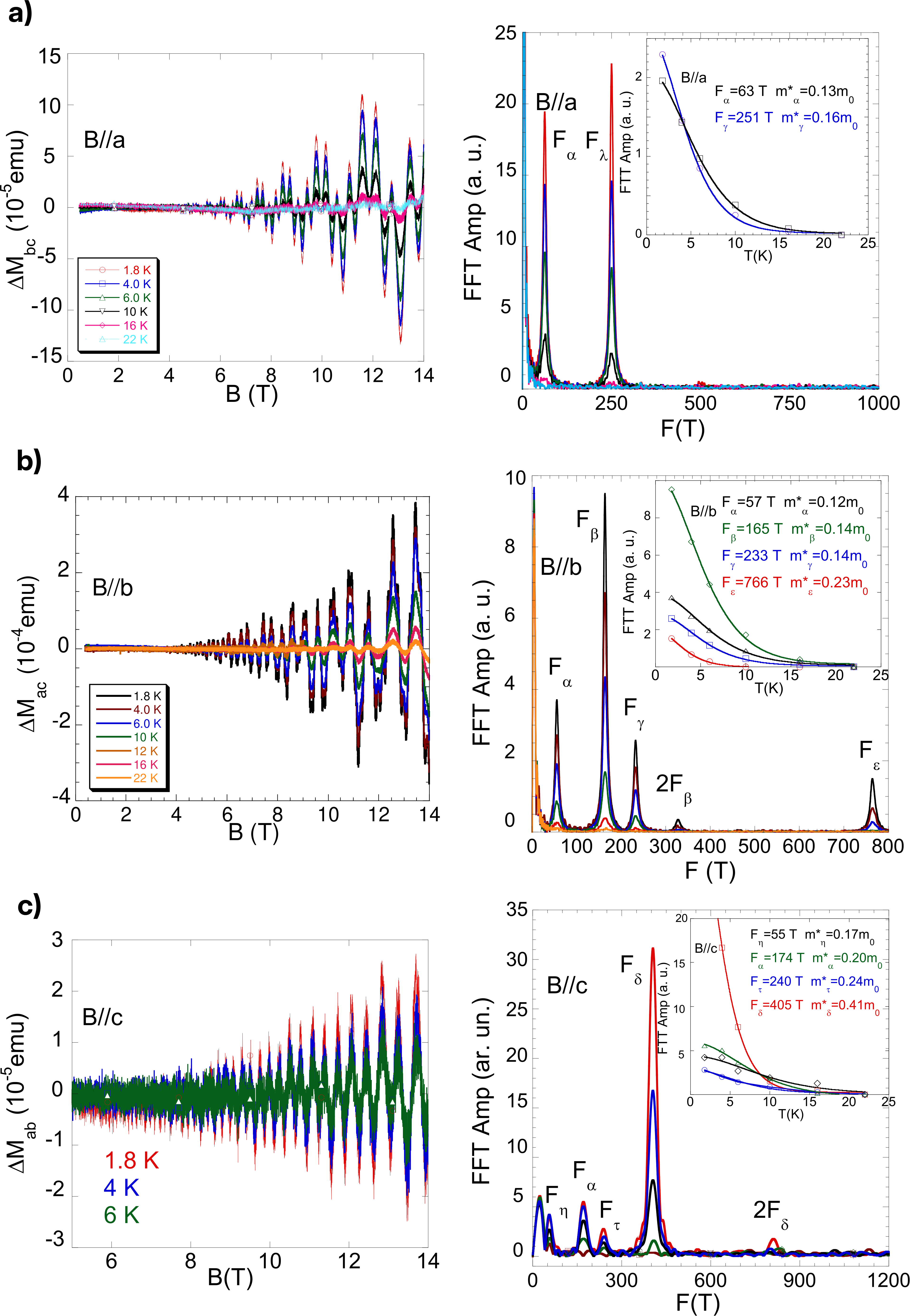}
	    \caption{Analysis of dHvA oscillations with a) \textbf{B}$\parallel$\emph{a}, b) \textbf{B}$\parallel$\emph{b}, and c) \textbf{B}$\parallel$\emph{c}. Left: The oscillatory components of the isothermal magnetization $\Delta$M after subtracting the diamagnetic background at various temperatures. Right: The corresponding FFT amplitude of the spectra has been generated from data between \SI{4}{\tesla} and \SI{14}{\tesla}. The inset shows the fits of the temperature damping factor $R_T$ in the LK formula to the FFT amplitudes.}
        \label{backgroundsubtr+FFT}
\end{figure}

\begin{table*}[ht]
\caption{\label{overview_magnetization_table} Extracted properties from de Haas van Alphen oscillations in the magnetization.}
\begin{ruledtabular}
\begin{tabular}{|c|c|c|c|c|c|c|c|c|}
\hline
& F (T) & $A_F$ (\SI{}{\angstrom^{-2}}) & $m^*$ ($m_e$) & $T_D$ (K) & $\tau_q$ ($10^{-14}$s) & $\mu_q\ $(m\textsuperscript{2}/Vs) & $l$ (nm)& $\phi$ ($\pi$) \\
\hline
\multicolumn{9}{c}{} \\
\multicolumn{9}{c}{\textbf{B}$\parallel$\emph{b}} \\
\hline
$F_\alpha$     & 56.8(3) & 0.00542(3) & 0.12(1) & 15.5(3) & 7.8(2) & 0.11(1) & 31(3) & 0.64(5) \\
$F_\beta$      & 164.7(3) & 0.01572(3) & 0.14(1) & 12.6(3) & 9.6(2) & 0.121(9) & 56(4) & 1.14(4) \\
$F_\gamma$     & 233.1(3) & 0.02225(3) & 0.14(2) & 12.5(5) & 9.7(4) & 0.12(2) & 70(10) & 0.98(3) \\
$F_\varepsilon$& 766.0(5) & 0.07312(5) & 0.23(2) & 12.3(5) & 9.9(4) & 0.076(7) & 76(7) & 0.81(4) \\
\hline
\multicolumn{9}{c}{} \\
\multicolumn{9}{c}{\textbf{B}$\parallel$\emph{c}} \\
\hline
$F_\eta$ & 55.3(3) & 0.00528(3) & 0.17(2) & 28.6(3) & 4.25(4) & 0.044(5) & 12(1) &  -  \\
$F_\alpha$  & 173.8(4) & 0.01659(4) & 0.20(2) & 13.4(3) & 9.1(2) & 0.080(8) & 38(4) & 1.03(3) \\
$F_\tau$ & 240.0(4) & 0.02291(4) & 0.24(3) & 13.6(4) & 8.9(3) & 0.066(8) & 37(5) &  -  \\
$F_\delta$ & 405.2(4) & 0.03868(4) & 0.41(2) & 4.7(3) & 26(2) & 0.111(9) & 81(7) & 1.30(3) \\
\hline
\multicolumn{9}{c}{} \\
\multicolumn{9}{c}{\textbf{B}$\parallel$\emph{a}} \\
\hline
$F_\alpha$ & 63.0(3) & 0.00601(3) & 0.13(1) & 15.8(3) & 7.7(1) & 0.104(8) & 30(2) & 1.10(5) \\
$F_\lambda$ & 251.0(4) & 0.02396(4) & 0.16(1) & 10.8(3) & 11.3(3) & 0.124(8) & 71(5) & 1.01(5) \\
\hline
\multicolumn{9}{c}{} \\
\multicolumn{9}{c}{\textbf{B}$\parallel$\emph{a} $\rightarrow$ \textbf{B}$\parallel$\emph{c} $\quad (\theta = \SI{15}{\degree})$} \\
\hline
$F_\alpha$ & 64.0(2) & 0.00611(2) & 0.19(1) & 12.9(5) & 9.4(4) & 0.087(6) & 25(2) & 0.62(3) \\
$F_\lambda$ & 254.0(4) & 0.02425(4) & 0.24(1) & 8.7(3) & 14.0(5) & 0.102(6) & 59(3) & 0.95(4) \\
\hline
\multicolumn{9}{c}{} \\
\multicolumn{9}{c}{\textbf{B}$\parallel$\emph{a} $\rightarrow$ \textbf{B}$\parallel$\emph{c} $\quad (\theta = \SI{35}{\degree})$} \\
\hline
$F_\alpha$ & 73.0(2) & 0.00697(2) & 0.18(1) & 12.4(4) & 9.8(3) & 0.096(6) & 30(2) & 0.61(3) \\
$F_\lambda$ & 286.0(4) & 0.02730(4) & 0.26(1) & 12.5(4) & 9.7(3) & 0.066(3) & 40(2) & 1.28(5) \\
\hline
\end{tabular}
\end{ruledtabular}
\end{table*}

The raw data for the magnetization measurements along all three crystallographic axes at different temperatures is given in \appref{raw_data_magnetization}. The diamagnetic background in this data was subtracted with a polynomial, and the resulting oscillatory part was fast Fourier transformed as illustrated in \autoref{backgroundsubtr+FFT}. From the FFT amplitudes, the effective mass $m^*$ can be extracted using the temperature-damping term $R_T$ as shown in the insets on the right. With the fitted effective mass as a known parameter, we have further fitted the oscillation pattern at \SI{1.8}{\kelvin} to the LK formula, from which the Dingle temperature $T_D$, quantum mobility $\mu_q$, and a phase shift $\phi$ of the Fermi pockets are extracted and listed in \autoref{overview_magnetization_table}. Only two orbits were considered in each fit at a time. After subtracting the corresponding oscillation pattern, we performed another two-band LK formula fit to the residual data (see \appref{analysis_magnetization}). Note, that for \textbf{B}$\parallel$\emph{c} the amplitude was too low to obtain reliable values for the phase shift $\phi$ for $F_{\eta}$ and $F_{\tau}$. The fitted Dingle temperature T$_D$, the quantum relaxation time $\tau_q$ and quantum mobility $\mu_q$ for all frequencies are also included in \autoref{overview_magnetization_table}.

\begin{figure}[hb]
        \includegraphics[width=0.49\columnwidth]{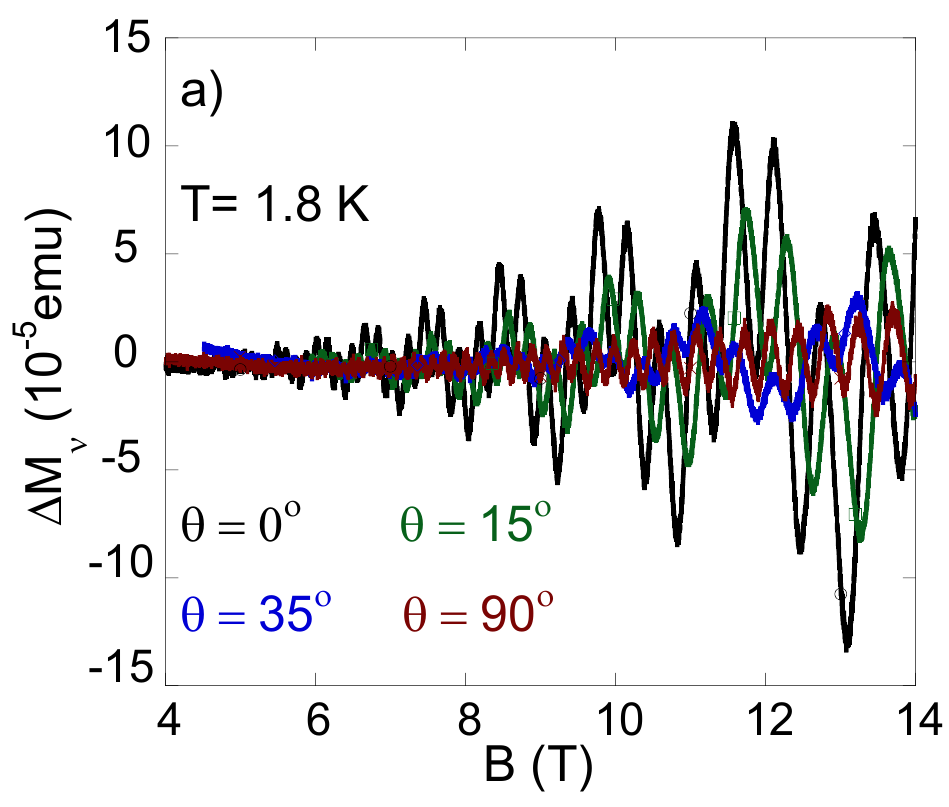}
        \includegraphics[width=0.49\columnwidth]{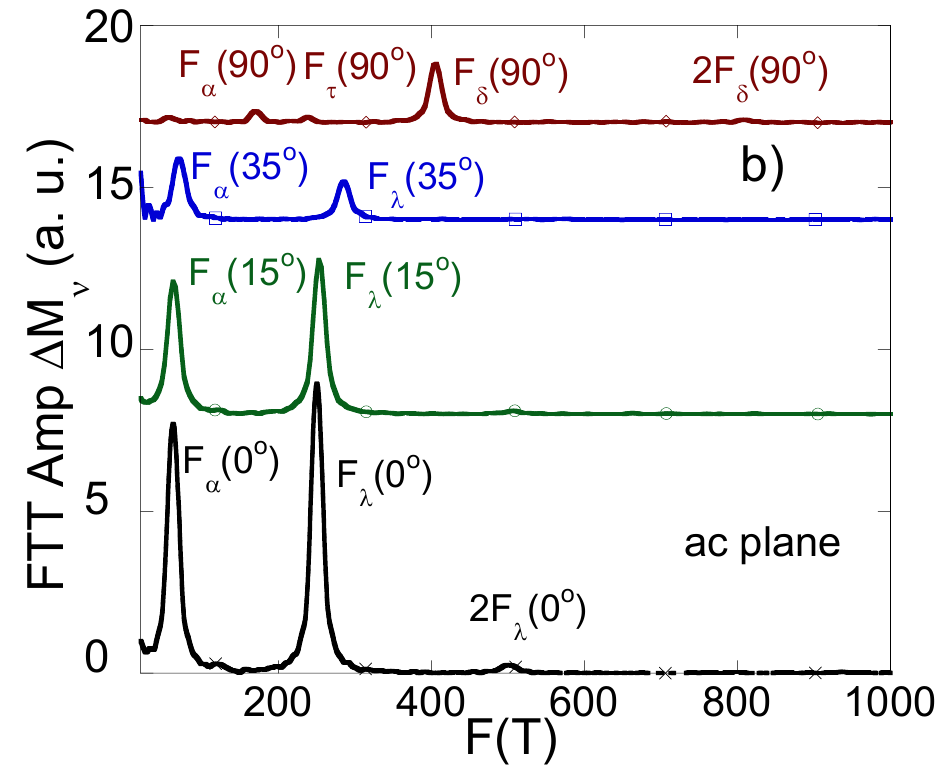}
    \caption{a) The magnetic field-dependent oscillatory components of the magnetization after subtracting the linear diamagnetic background at different angles in the \emph{a}-\emph{c} plane at T = \SI{1.8}{\kelvin}. b) The corresponding FFT spectra for different angles. The angle $\theta$ in the \emph{a}-\emph{c} plane is defined as the angle between the applied magnetic field and the \emph{a}-axis.}
    \label{BIIa-c_sweep}
\end{figure}

To probe the morphology of the Fermi surface, we furthermore carried out angle-dependent quantum oscillation measurements in the magnetization with the applied magnetic field in the crystallographic \emph{a}-\emph{c} plane. The background subtracted raw data for four angles in that plane including its respective FFT is shown \autoref{BIIa-c_sweep}.

\begin{figure*}[ht]
        \includegraphics[width=0.98\textwidth]{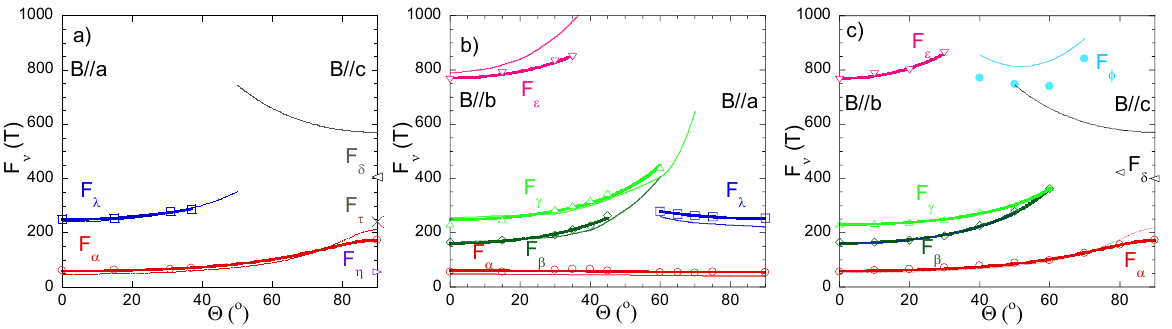}
	\caption {The angular dependence of the oscillation frequencies. The $F_{\alpha}$ branch was fitted with the ellipsoid model, while all other branches were fitted with the $F_{2D}/\cos(\theta)+F_{3D}$ model described in the main text and illustrated by thick lines. DFT calculations are included via thin lines. a) Magnetization in the \emph{a}-\emph{c} plane. b) Magnetic torque in the \emph{a}-\emph{b} plane. c) Magnetic torque in the \emph{b}-\emph{c} plane.}
    \label{BIIa-b_sweep}
\end{figure*}

With increasing angle $\theta$ from \SI{0}{\degree} to \SI{90}{\degree}, the oscillation components are suppressed gradually. It can be seen from \autoref{BIIa-b_sweep} a) that with increasing the angle $\theta$, the oscillation frequency $F_\alpha$ changes only slightly from \SI{63}{\tesla} ($\theta = \SI{0}{\degree}$) to \SI{73}{\tesla} ($\theta = \SI{35}{\degree}$), before increasing steeply to \SI{173.8}{\tesla} at $\theta = \SI{90}{\degree}$, indicating a 3D elliptical Fermi pocket. This motivates fitting the angle dependence of the $F_\alpha$ pocket with the 3D ellipsoid model (thick red line) described in \cite{julian2012numerical}. On the other hand, the frequency $F_\lambda$ increases from \SI{251}{\tesla} ($\theta = \SI{0}{\degree}$) to \SI{286}{\tesla} ($\theta = \SI{35}{\degree}$), and disappears for the angles larger than ($\theta = \SI{35}{\degree}$). This motivates the model of a 2-dimensional Fermi sheet (FS), and hence we fit the data with $F_{2D}/\cos(\theta)+F_{3D}$ \cite{hu2017nearly} (thick blue line), which includes both 2D and isotropic 3D FS components. We find the ratio $F_{2D}/F_{3D}\approx 1.65$, and conclude that the dimensionality of the FS associated with $F_\lambda$ is a mixture of 2D and 3D character where the former dominates. We also include the results from DFT calculations in \autoref{BIIa-b_sweep} a) (thin lines) which seem to agree well with the experimental findings after shifting the Fermi energy by \SI{20}{\milli\electronvolt}. Particularly bands 2 and 5 (see \autoref{BZ_bandstructure}) can explain the presence of the branches $F_{\lambda}$ and $F_{\alpha}$, respectively, while band 1 seems to overestimate $F_{\delta}$. The frequencies $F_\tau$=\SI{240}{\tesla} and $F_\eta$=\SI{55.3}{\tesla} for \textbf{B}$\parallel$\emph{c} are two new frequencies, which have not been reported in the literature thus far. With the strong anisotropy of the crystal, the oscillatory signal is the weakest along the \emph{c} direction, and more measurements are necessary to conclusively confirm the existence of these two additional frequencies. Effective masses and other quantities were extracted for data points in the \emph{a}-\emph{c} plane and are included in \autoref{overview_magnetization_table}. Note that an additional data point at around \SI{30}{\degree} is included in \autoref{BIIa-b_sweep} a), but due to the lack of a temperature-dependent study, only its frequency could be determined.

\subsection{Magnetic Torque}
To obtain a better understanding of the morphology of the Fermi surface, we carried out angle-dependent quantum oscillation measurements in the magnetic torque with the magnetic field in the \emph{a}-\emph{b} plane up to \SI{9}{\tesla} and in the \emph{b}-\emph{c} plane up to \SI{13}{\tesla}. An analysis is shown in \appref{raw_data_magnetic_torque}, and the results are summarized \autoref{BIIa-b_sweep} b) and \autoref{BIIa-b_sweep} c). Again, the oscillation frequency $F_\alpha$ can be well fitted with the 3D ellipsoid model. In the \emph{a}-\emph{b} plane, $F_\lambda$  decreases from \SI{276}{\tesla} ($\theta = \SI{60}{\degree}$) to \SI{251}{\tesla} ($\theta = \SI{90}{\degree}$), while disappearing for the angles smaller than $\theta = \SI{60}{\degree}$. This dependence is fitted by $F_{2D}/\cos(\theta)+F_{3D}$, and results in a ratio of $F_{2D}/F_{3D}\approx 1.9$. With increasing the angle $\theta$, the highest frequency branch $F_\varepsilon$ continues to increase until it disappears above $\theta =$\SI{35}{\degree}. For $F_\beta$, $F_\gamma$ and $F_\varepsilon$ we find ratios of $F_{2D}/F_{3D}\approx 16.5$, $F_{2D}/F_{3D}\approx 4.1$ and $F_{2D}/F_{3D}\approx 0.9$, respectively. Comparing these results to the rotation study in magnetic torque in the \emph{a}-\emph{b} plane performed by Z. Chen et al. \cite{chen21}, we have observed agreement for the low frequency branch $F_\alpha$, both in magnitude and the angle-dependence. Z. Chen et al. \cite{chen21} also report oscillation data for the electron-branch $F_\lambda$  and lower hole-branch $F_\beta$, although they mistakenly conclude that these frequencies must arise from the same Fermi surface. The upper hole-branch $F_\gamma$ was not reported at all. Our DFT analysis shows that the branches $F_\beta$ and $F_\gamma$ arise from band 1, while $F_\lambda$ arises from band 2 as noted before. Furthermore, Z. Chen et al. \cite{chen21} also report the hole-branch $F_\varepsilon$ and an additional branch with similar angular dependence with a 150 T larger frequency. However, both results disagree with the authors' own Shubnikov-de Haas study, which finds only one instead of two frequencies at a value more consistent with our measurements. We find that $F_\varepsilon$ originates from the Fermi sheet in band 2.

\begin{table*}[ht]
\caption{\label{overview_resistivity_table} Extracted properties from Shubnikov de Haas oscillations in the resistivity with \textbf{B}$\parallel$\emph{b} and \textbf{I}$\parallel$\emph{a}.}
\begin{ruledtabular}
\begin{tabular}{|c|c|c|c|c|c|c|c|c|}
\hline
& F (T) & $A_F$ (\SI{}{\angstrom^{-2}}) & $m^*$ ($m_e$) & $T_D$ (K) & $\tau_q$ ($10^{-14}$s) & $\mu_q\ $(m$\textsuperscript{2}$/Vs)&  $l$ (nm)& $\phi$ ($\pi$)  \\
\hline
$F_\alpha$     & 53.0(3) & 0.00506(3) & 0.14(2) & 12.9(4) & 9.4(3) & 0.12(2) & 31(5) & 1.03(3) \\
$F_\beta$      & 163.0(4) & 0.01556(4) & 0.23(1) & 14.7(3) & 8.3(2) & 0.063(3) & 29(1) & 1.05(4) \\
$F_\gamma$     & 233(3) & 0.0222(3) & 0.23(3) &  -   &  -  &   -     & - & -  \\
$F_\varepsilon$& 766(5) &  -     &  -   &  -   &  -  &   -     & - & -  \\
\hline
\end{tabular}
\end{ruledtabular}
\end{table*}

The angle dependence of the oscillation frequencies from the torque measurements in the \emph{b}-\emph{c} plane are shown in \autoref{BIIa-b_sweep} c). The frequency branches show similar behavior as in the \emph{a}-\emph{b} plane with ratios of $F_{2D}/F_{3D}\approx 17.2$ for the $F_\beta$ pocket, $F_{2D}/F_{3D}\approx 1.4$ for the $F_\gamma$ pocket and $F_{2D}/F_{3D}\approx 3.5$ for the $F_\varepsilon$ pocket. No attempt was made to fit $F_\phi$, but regarding DFT calculations we find that it most likely arises from band 1.

\subsection{Magnetoresistance}

\begin{figure}[ht]
        \includegraphics[width=0.45\columnwidth]{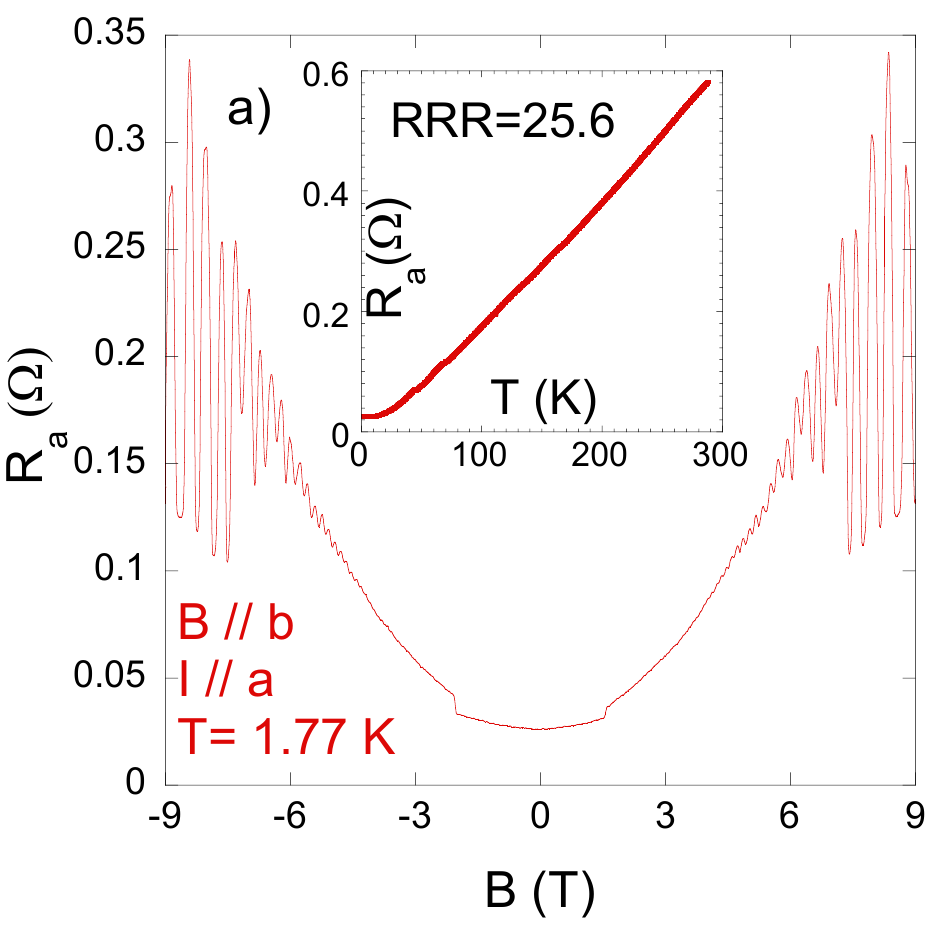}
        \includegraphics[width=0.53\columnwidth]{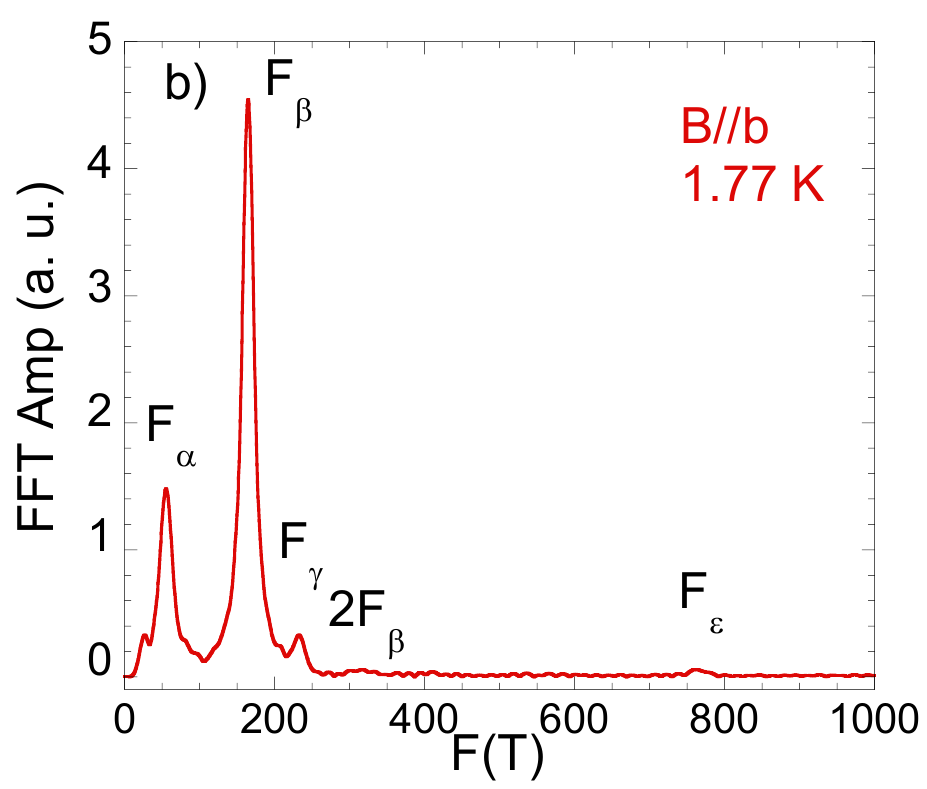}
	\caption{a) Magnetic field dependent resistance R$_a$ with the current applied along the \emph{a}-axis, and the magnetic field parallel to the \emph{b}-axis. The inset shows the temperature dependence of the resistance R$_a$ at \SI{0}{\tesla}. b) The FFT spectrum of $\Delta$R$_a$ generated in the magnetic field range from 3 to \SI{9}{\tesla}.}
        \label{resistivity}
\end{figure}

\autoref{resistivity} a) shows the magnetoresistance data with the magnetic field parallel to the \emph{b}-axis and the applied current parallel to the \emph{a}-axis. The oscillations in the magnetoresistance shown here are significantly more pronounced than what has been reported in recent works \cite{xu2020anisotropic,chen21}, in which quantum oscillations in the magnetoresistance of TaNiTe$_5$ were only observed at magnetic fields of up to 50T. The strong quantum oscillations shown here indicate a possible new topological phenomenon, however a model based on classical phenomena can fully explain the origin of this amplification of quantum oscillations as described in more detail in \cite{daschner2025mechanical}.

The inset to \autoref{resistivity} a) shows the temperature dependence of the resistance R$_a$ displaying a typical metallic behavior. The resistance follows a $T$-linear dependence from 280K to 50K and then a cross-over to a quadratic $T$-dependence indicating Fermi liquid behavior below 50K. The residual resistance ratio is measured as RRR = 25.6. Moreover, the quasi-1D transport behavior with a resistance ratio of $\rho_a$:$\rho_c = $1:4 at \SI{2}{\kelvin} has been obtained and is consistent with previous reports \cite{xu2020anisotropic,chen21}. Using this data, we performed a fast Fourier transform in the magnetic field range between \SI{3}{\tesla} and \SI{9}{\tesla} as shown in \autoref{resistivity} b).

Similar to the magnetization data, magnetoresistance data in the temperature range \SI{1.77}{\kelvin} - \SI{20}{\kelvin} allows us to extract the effective mass, and other parameters (see \appref{raw_data_magnetoresistance}). Given $m^*$, the Dingle temperature $T_D$, quantum mobility $\mu_q$, and the phase shift $\phi_B$ of the Fermi pockets are extracted at T = \SI{1.77}{\kelvin}, and summarized \autoref{overview_resistivity_table}. The resolution of the $\gamma$ and $\varepsilon$ pockets is too low to extract most parameters other than the frequencies, and the effective mass for the $\gamma$ pocket.

\section{Conclusions}
We performed high-resolution measurements of de Haas - van Alphen and Shubnikov - de Haas quantum oscillations in layered TaNiTe$_5$ along all three crystallographic axes and the planes in between. The obtained fundamental frequencies in the FFT spectra resulting from dHvA measurements are consistent with those found in SdH measurements when the magnetic field is applied parallel to the \emph{b}-axis. Furthermore, quantum oscillation measurements in the magnetization, magnetic torque and the magnetoresistance are consistent with previously reported values in the literature and density functional theory calculations performed here. Our rotation studies allow us to map out the morphology of the Fermi surface in this material, and we discovered new fundamental frequencies that had been absent from the literature thus far.

\begin{acknowledgments}
This work was supported by the Croatian Science Foundation under the project IP 2018 01 8912 and by the EPSRC of the UK under grant EP/X011992/1. This work was also supported by the Henry Royce Institute for advanced materials through the Equipment Access Scheme enabling access to the Magnetic Property Measurement System (MPMS) at Cambridge; Cambridge Royce facilities grant EP/P024947/1 and Sir Henry Royce Institute - recurrent grant EP/R00661X/1)
\end{acknowledgments}

\bibliography{Main}

\begin{thebibliography}{37}%
\makeatletter
\providecommand \@ifxundefined [1]{%
 \@ifx{#1\undefined}
}%
\providecommand \@ifnum [1]{%
 \ifnum #1\expandafter \@firstoftwo
 \else \expandafter \@secondoftwo
 \fi
}%
\providecommand \@ifx [1]{%
 \ifx #1\expandafter \@firstoftwo
 \else \expandafter \@secondoftwo
 \fi
}%
\providecommand \natexlab [1]{#1}%
\providecommand \enquote  [1]{``#1''}%
\providecommand \bibnamefont  [1]{#1}%
\providecommand \bibfnamefont [1]{#1}%
\providecommand \citenamefont [1]{#1}%
\providecommand \href@noop [0]{\@secondoftwo}%
\providecommand \href [0]{\begingroup \@sanitize@url \@href}%
\providecommand \@href[1]{\@@startlink{#1}\@@href}%
\providecommand \@@href[1]{\endgroup#1\@@endlink}%
\providecommand \@sanitize@url [0]{\catcode `\\12\catcode `\$12\catcode `\&12\catcode `\#12\catcode `\^12\catcode `\_12\catcode `\%12\relax}%
\providecommand \@@startlink[1]{}%
\providecommand \@@endlink[0]{}%
\providecommand \url  [0]{\begingroup\@sanitize@url \@url }%
\providecommand \@url [1]{\endgroup\@href {#1}{\urlprefix }}%
\providecommand \urlprefix  [0]{URL }%
\providecommand \Eprint [0]{\href }%
\providecommand \doibase [0]{https://doi.org/}%
\providecommand \selectlanguage [0]{\@gobble}%
\providecommand \bibinfo  [0]{\@secondoftwo}%
\providecommand \bibfield  [0]{\@secondoftwo}%
\providecommand \translation [1]{[#1]}%
\providecommand \BibitemOpen [0]{}%
\providecommand \bibitemStop [0]{}%
\providecommand \bibitemNoStop [0]{.\EOS\space}%
\providecommand \EOS [0]{\spacefactor3000\relax}%
\providecommand \BibitemShut  [1]{\csname bibitem#1\endcsname}%
\let\auto@bib@innerbib\@empty
\bibitem [{\citenamefont {Young}\ and\ \citenamefont {Kane}(2015)}]{young2015dirac}%
  \BibitemOpen
  \bibfield  {author} {\bibinfo {author} {\bibfnamefont {S.~M.}\ \bibnamefont {Young}}\ and\ \bibinfo {author} {\bibfnamefont {C.~L.}\ \bibnamefont {Kane}},\ }\href@noop {} {\bibfield  {journal} {\bibinfo  {journal} {Physical review letters}\ }\textbf {\bibinfo {volume} {115}},\ \bibinfo {pages} {126803} (\bibinfo {year} {2015})}\BibitemShut {NoStop}%
\bibitem [{\citenamefont {Yan}\ \emph {et~al.}(2017)\citenamefont {Yan}, \citenamefont {Huang}, \citenamefont {Zhang}, \citenamefont {Wang}, \citenamefont {Yao}, \citenamefont {Deng}, \citenamefont {Wan}, \citenamefont {Zhang}, \citenamefont {Arita}, \citenamefont {Yang} \emph {et~al.}}]{yan17}%
  \BibitemOpen
  \bibfield  {author} {\bibinfo {author} {\bibfnamefont {M.}~\bibnamefont {Yan}}, \bibinfo {author} {\bibfnamefont {H.}~\bibnamefont {Huang}}, \bibinfo {author} {\bibfnamefont {K.}~\bibnamefont {Zhang}}, \bibinfo {author} {\bibfnamefont {E.}~\bibnamefont {Wang}}, \bibinfo {author} {\bibfnamefont {W.}~\bibnamefont {Yao}}, \bibinfo {author} {\bibfnamefont {K.}~\bibnamefont {Deng}}, \bibinfo {author} {\bibfnamefont {G.}~\bibnamefont {Wan}}, \bibinfo {author} {\bibfnamefont {H.}~\bibnamefont {Zhang}}, \bibinfo {author} {\bibfnamefont {M.}~\bibnamefont {Arita}}, \bibinfo {author} {\bibfnamefont {H.}~\bibnamefont {Yang}}, \emph {et~al.},\ }\href@noop {} {\bibfield  {journal} {\bibinfo  {journal} {Nature communications}\ }\textbf {\bibinfo {volume} {8}},\ \bibinfo {pages} {257} (\bibinfo {year} {2017})}\BibitemShut {NoStop}%
\bibitem [{\citenamefont {Xiong}\ \emph {et~al.}(2015)\citenamefont {Xiong}, \citenamefont {Kushwaha}, \citenamefont {Liang}, \citenamefont {Krizan}, \citenamefont {Hirschberger}, \citenamefont {Wang}, \citenamefont {Cava},\ and\ \citenamefont {Ong}}]{xiong15}%
  \BibitemOpen
  \bibfield  {author} {\bibinfo {author} {\bibfnamefont {J.}~\bibnamefont {Xiong}}, \bibinfo {author} {\bibfnamefont {S.~K.}\ \bibnamefont {Kushwaha}}, \bibinfo {author} {\bibfnamefont {T.}~\bibnamefont {Liang}}, \bibinfo {author} {\bibfnamefont {J.~W.}\ \bibnamefont {Krizan}}, \bibinfo {author} {\bibfnamefont {M.}~\bibnamefont {Hirschberger}}, \bibinfo {author} {\bibfnamefont {W.}~\bibnamefont {Wang}}, \bibinfo {author} {\bibfnamefont {R.~J.}\ \bibnamefont {Cava}},\ and\ \bibinfo {author} {\bibfnamefont {N.~P.}\ \bibnamefont {Ong}},\ }\href@noop {} {\bibfield  {journal} {\bibinfo  {journal} {Science}\ }\textbf {\bibinfo {volume} {350}},\ \bibinfo {pages} {413} (\bibinfo {year} {2015})}\BibitemShut {NoStop}%
\bibitem [{\citenamefont {Borisenko}\ \emph {et~al.}(2014)\citenamefont {Borisenko}, \citenamefont {Gibson}, \citenamefont {Evtushinsky}, \citenamefont {Zabolotnyy}, \citenamefont {Buchner},\ and\ \citenamefont {Cava}}]{borisenko14}%
  \BibitemOpen
  \bibfield  {author} {\bibinfo {author} {\bibfnamefont {S.}~\bibnamefont {Borisenko}}, \bibinfo {author} {\bibfnamefont {Q.}~\bibnamefont {Gibson}}, \bibinfo {author} {\bibfnamefont {D.}~\bibnamefont {Evtushinsky}}, \bibinfo {author} {\bibfnamefont {V.}~\bibnamefont {Zabolotnyy}}, \bibinfo {author} {\bibfnamefont {B.}~\bibnamefont {Buchner}},\ and\ \bibinfo {author} {\bibfnamefont {R.~J.}\ \bibnamefont {Cava}},\ }\href@noop {} {\bibfield  {journal} {\bibinfo  {journal} {Phys. Rev. Lett.}\ }\textbf {\bibinfo {volume} {113}},\ \bibinfo {pages} {027603} (\bibinfo {year} {2014})}\BibitemShut {NoStop}%
\bibitem [{\citenamefont {Huang}\ \emph {et~al.}(2015)\citenamefont {Huang}, \citenamefont {Xu}, \citenamefont {Belopolski}, \citenamefont {Lee}, \citenamefont {Chang}, \citenamefont {Wang}, \citenamefont {Alidoust}, \citenamefont {Bian}, \citenamefont {Neupane}, \citenamefont {Zhang}, \citenamefont {Jia}, \citenamefont {Bansil}, \citenamefont {Lin},\ and\ \citenamefont {Hasan}}]{huang15}%
  \BibitemOpen
  \bibfield  {author} {\bibinfo {author} {\bibfnamefont {S.~M.}\ \bibnamefont {Huang}}, \bibinfo {author} {\bibfnamefont {S.~Y.}\ \bibnamefont {Xu}}, \bibinfo {author} {\bibfnamefont {I.}~\bibnamefont {Belopolski}}, \bibinfo {author} {\bibfnamefont {C.~C.}\ \bibnamefont {Lee}}, \bibinfo {author} {\bibfnamefont {G.}~\bibnamefont {Chang}}, \bibinfo {author} {\bibfnamefont {B.}~\bibnamefont {Wang}}, \bibinfo {author} {\bibfnamefont {N.}~\bibnamefont {Alidoust}}, \bibinfo {author} {\bibfnamefont {G.}~\bibnamefont {Bian}}, \bibinfo {author} {\bibfnamefont {M.}~\bibnamefont {Neupane}}, \bibinfo {author} {\bibfnamefont {C.}~\bibnamefont {Zhang}}, \bibinfo {author} {\bibfnamefont {S.}~\bibnamefont {Jia}}, \bibinfo {author} {\bibfnamefont {A.}~\bibnamefont {Bansil}}, \bibinfo {author} {\bibfnamefont {H.}~\bibnamefont {Lin}},\ and\ \bibinfo {author} {\bibfnamefont {M.~Z.}\ \bibnamefont {Hasan}},\ }\href@noop {} {\bibfield  {journal} {\bibinfo  {journal} {Nat. Commun.}\ }\textbf {\bibinfo {volume} {6}},\ \bibinfo {pages} {7373} (\bibinfo {year} {2015})}\BibitemShut {NoStop}%
\bibitem [{\citenamefont {Soluyanov}\ \emph {et~al.}(2015)\citenamefont {Soluyanov}, \citenamefont {Gresch}, \citenamefont {Wang}, \citenamefont {Wu}, \citenamefont {Troyer}, \citenamefont {Dai},\ and\ \citenamefont {Bernevig}}]{soluyanov15}%
  \BibitemOpen
  \bibfield  {author} {\bibinfo {author} {\bibfnamefont {A.~A.}\ \bibnamefont {Soluyanov}}, \bibinfo {author} {\bibfnamefont {D.}~\bibnamefont {Gresch}}, \bibinfo {author} {\bibfnamefont {Z.}~\bibnamefont {Wang}}, \bibinfo {author} {\bibfnamefont {Q.}~\bibnamefont {Wu}}, \bibinfo {author} {\bibfnamefont {M.}~\bibnamefont {Troyer}}, \bibinfo {author} {\bibfnamefont {X.}~\bibnamefont {Dai}},\ and\ \bibinfo {author} {\bibfnamefont {B.~A.}\ \bibnamefont {Bernevig}},\ }\href@noop {} {\bibfield  {journal} {\bibinfo  {journal} {Nature (London)}\ }\textbf {\bibinfo {volume} {527}},\ \bibinfo {pages} {495} (\bibinfo {year} {2015})}\BibitemShut {NoStop}%
\bibitem [{\citenamefont {Weng}\ \emph {et~al.}(2015)\citenamefont {Weng}, \citenamefont {Fang}, \citenamefont {Fang}, \citenamefont {Bernevig},\ and\ \citenamefont {Dai}}]{weng15}%
  \BibitemOpen
  \bibfield  {author} {\bibinfo {author} {\bibfnamefont {H.}~\bibnamefont {Weng}}, \bibinfo {author} {\bibfnamefont {C.}~\bibnamefont {Fang}}, \bibinfo {author} {\bibfnamefont {Z.}~\bibnamefont {Fang}}, \bibinfo {author} {\bibfnamefont {B.~A.}\ \bibnamefont {Bernevig}},\ and\ \bibinfo {author} {\bibfnamefont {X.}~\bibnamefont {Dai}},\ }\href@noop {} {\bibfield  {journal} {\bibinfo  {journal} {Phys. Rev. X}\ }\textbf {\bibinfo {volume} {5}},\ \bibinfo {pages} {011029} (\bibinfo {year} {2015})}\BibitemShut {NoStop}%
\bibitem [{\citenamefont {Lv}\ \emph {et~al.}(2015)\citenamefont {Lv}, \citenamefont {Xu}, \citenamefont {Weng}, \citenamefont {Ma}, \citenamefont {Richard}, \citenamefont {Huang}, \citenamefont {Zhao}, \citenamefont {Chen}, \citenamefont {Matt}, \citenamefont {Bisti}, \citenamefont {Strocov}, \citenamefont {Mesot}, \citenamefont {Fang}, \citenamefont {Dai}, \citenamefont {Qian}, \citenamefont {Shi},\ and\ \citenamefont {Ding}}]{lv15}%
  \BibitemOpen
  \bibfield  {author} {\bibinfo {author} {\bibfnamefont {B.~Q.}\ \bibnamefont {Lv}}, \bibinfo {author} {\bibfnamefont {N.}~\bibnamefont {Xu}}, \bibinfo {author} {\bibfnamefont {H.~M.}\ \bibnamefont {Weng}}, \bibinfo {author} {\bibfnamefont {J.~Z.}\ \bibnamefont {Ma}}, \bibinfo {author} {\bibfnamefont {P.}~\bibnamefont {Richard}}, \bibinfo {author} {\bibfnamefont {X.~C.}\ \bibnamefont {Huang}}, \bibinfo {author} {\bibfnamefont {L.~X.}\ \bibnamefont {Zhao}}, \bibinfo {author} {\bibfnamefont {G.~F.}\ \bibnamefont {Chen}}, \bibinfo {author} {\bibfnamefont {C.~E.}\ \bibnamefont {Matt}}, \bibinfo {author} {\bibfnamefont {F.}~\bibnamefont {Bisti}}, \bibinfo {author} {\bibfnamefont {V.~N.}\ \bibnamefont {Strocov}}, \bibinfo {author} {\bibfnamefont {J.}~\bibnamefont {Mesot}}, \bibinfo {author} {\bibfnamefont {Z.}~\bibnamefont {Fang}}, \bibinfo {author} {\bibfnamefont {X.}~\bibnamefont {Dai}}, \bibinfo {author} {\bibfnamefont {T.}~\bibnamefont {Qian}}, \bibinfo {author} {\bibfnamefont {M.}~\bibnamefont {Shi}},\ and\ \bibinfo {author} {\bibfnamefont {H.}~\bibnamefont {Ding}},\ }\href@noop {} {\bibfield  {journal} {\bibinfo  {journal} {Nat. Phys.}\ }\textbf {\bibinfo {volume} {11}},\ \bibinfo {pages} {724} (\bibinfo {year} {2015})}\BibitemShut {NoStop}%
\bibitem [{\citenamefont {Xu}\ \emph {et~al.}(2015{\natexlab{a}})\citenamefont {Xu}, \citenamefont {Alidoust}, \citenamefont {Belopolski}, \citenamefont {Yuan}, \citenamefont {Bian}, \citenamefont {Chang}, \citenamefont {Zheng}, \citenamefont {Strocov}, \citenamefont {Sanchez}, \citenamefont {Chang} \emph {et~al.}}]{xu15a}%
  \BibitemOpen
  \bibfield  {author} {\bibinfo {author} {\bibfnamefont {S.-Y.}\ \bibnamefont {Xu}}, \bibinfo {author} {\bibfnamefont {N.}~\bibnamefont {Alidoust}}, \bibinfo {author} {\bibfnamefont {I.}~\bibnamefont {Belopolski}}, \bibinfo {author} {\bibfnamefont {Z.}~\bibnamefont {Yuan}}, \bibinfo {author} {\bibfnamefont {G.}~\bibnamefont {Bian}}, \bibinfo {author} {\bibfnamefont {T.-R.}\ \bibnamefont {Chang}}, \bibinfo {author} {\bibfnamefont {H.}~\bibnamefont {Zheng}}, \bibinfo {author} {\bibfnamefont {V.~N.}\ \bibnamefont {Strocov}}, \bibinfo {author} {\bibfnamefont {D.~S.}\ \bibnamefont {Sanchez}}, \bibinfo {author} {\bibfnamefont {G.}~\bibnamefont {Chang}}, \emph {et~al.},\ }\href@noop {} {\bibfield  {journal} {\bibinfo  {journal} {Nature Physics}\ }\textbf {\bibinfo {volume} {11}},\ \bibinfo {pages} {748} (\bibinfo {year} {2015}{\natexlab{a}})}\BibitemShut {NoStop}%
\bibitem [{\citenamefont {Xu}\ \emph {et~al.}(2015{\natexlab{b}})\citenamefont {Xu}, \citenamefont {Belopolski}, \citenamefont {Alidoust}, \citenamefont {Neupane}, \citenamefont {Bian}, \citenamefont {Zhang}, \citenamefont {Sankar}, \citenamefont {Chang}, \citenamefont {Yuan}, \citenamefont {Lee} \emph {et~al.}}]{xu15b}%
  \BibitemOpen
  \bibfield  {author} {\bibinfo {author} {\bibfnamefont {S.-Y.}\ \bibnamefont {Xu}}, \bibinfo {author} {\bibfnamefont {I.}~\bibnamefont {Belopolski}}, \bibinfo {author} {\bibfnamefont {N.}~\bibnamefont {Alidoust}}, \bibinfo {author} {\bibfnamefont {M.}~\bibnamefont {Neupane}}, \bibinfo {author} {\bibfnamefont {G.}~\bibnamefont {Bian}}, \bibinfo {author} {\bibfnamefont {C.}~\bibnamefont {Zhang}}, \bibinfo {author} {\bibfnamefont {R.}~\bibnamefont {Sankar}}, \bibinfo {author} {\bibfnamefont {G.}~\bibnamefont {Chang}}, \bibinfo {author} {\bibfnamefont {Z.}~\bibnamefont {Yuan}}, \bibinfo {author} {\bibfnamefont {C.-C.}\ \bibnamefont {Lee}}, \emph {et~al.},\ }\href@noop {} {\bibfield  {journal} {\bibinfo  {journal} {Science}\ }\textbf {\bibinfo {volume} {349}},\ \bibinfo {pages} {613} (\bibinfo {year} {2015}{\natexlab{b}})}\BibitemShut {NoStop}%
\bibitem [{\citenamefont {Xu}\ \emph {et~al.}(2015{\natexlab{c}})\citenamefont {Xu}, \citenamefont {Liu}, \citenamefont {Kushwaha}, \citenamefont {Sankar}, \citenamefont {Krizan}, \citenamefont {Belopolski}, \citenamefont {Neupane}, \citenamefont {Bian}, \citenamefont {Alidoust}, \citenamefont {Chang}, \citenamefont {Jeng}, \citenamefont {Huang}, \citenamefont {Tsai}, \citenamefont {Lin}, \citenamefont {Shibayev}, \citenamefont {Chou}, \citenamefont {Cava},\ and\ \citenamefont {Hasan}}]{xu15c}%
  \BibitemOpen
  \bibfield  {author} {\bibinfo {author} {\bibfnamefont {S.-Y.}\ \bibnamefont {Xu}}, \bibinfo {author} {\bibfnamefont {C.}~\bibnamefont {Liu}}, \bibinfo {author} {\bibfnamefont {S.~K.}\ \bibnamefont {Kushwaha}}, \bibinfo {author} {\bibfnamefont {R.}~\bibnamefont {Sankar}}, \bibinfo {author} {\bibfnamefont {J.~W.}\ \bibnamefont {Krizan}}, \bibinfo {author} {\bibfnamefont {I.}~\bibnamefont {Belopolski}}, \bibinfo {author} {\bibfnamefont {M.}~\bibnamefont {Neupane}}, \bibinfo {author} {\bibfnamefont {G.}~\bibnamefont {Bian}}, \bibinfo {author} {\bibfnamefont {N.}~\bibnamefont {Alidoust}}, \bibinfo {author} {\bibfnamefont {T.-R.}\ \bibnamefont {Chang}}, \bibinfo {author} {\bibfnamefont {H.-T.}\ \bibnamefont {Jeng}}, \bibinfo {author} {\bibfnamefont {C.-Y.}\ \bibnamefont {Huang}}, \bibinfo {author} {\bibfnamefont {W.-F.}\ \bibnamefont {Tsai}}, \bibinfo {author} {\bibfnamefont {H.}~\bibnamefont {Lin}}, \bibinfo {author} {\bibfnamefont {P.~P.}\ \bibnamefont {Shibayev}}, \bibinfo {author} {\bibfnamefont {F.-C.}\ \bibnamefont {Chou}}, \bibinfo {author} {\bibfnamefont {R.~J.}\ \bibnamefont {Cava}},\ and\ \bibinfo {author} {\bibfnamefont {M.~Z.}\ \bibnamefont {Hasan}},\ }\href@noop {} {\bibfield  {journal} {\bibinfo  {journal} {Science}\ }\textbf {\bibinfo {volume} {347}},\ \bibinfo {pages} {294} (\bibinfo {year} {2015}{\natexlab{c}})}\BibitemShut {NoStop}%
\bibitem [{\citenamefont {Yang}\ \emph {et~al.}(2015)\citenamefont {Yang}, \citenamefont {Liu}, \citenamefont {Sun}, \citenamefont {Peng}, \citenamefont {Yang}, \citenamefont {Zhang}, \citenamefont {Zhou}, \citenamefont {Zhang}, \citenamefont {Guo}, \citenamefont {Rahn}, \citenamefont {Prabhakaran}, \citenamefont {Hussain}, \citenamefont {Mo}, \citenamefont {Felser}, \citenamefont {Yan},\ and\ \citenamefont {Chen}}]{yang15}%
  \BibitemOpen
  \bibfield  {author} {\bibinfo {author} {\bibfnamefont {L.~X.}\ \bibnamefont {Yang}}, \bibinfo {author} {\bibfnamefont {Z.~K.}\ \bibnamefont {Liu}}, \bibinfo {author} {\bibfnamefont {Y.}~\bibnamefont {Sun}}, \bibinfo {author} {\bibfnamefont {H.}~\bibnamefont {Peng}}, \bibinfo {author} {\bibfnamefont {H.~F.}\ \bibnamefont {Yang}}, \bibinfo {author} {\bibfnamefont {T.}~\bibnamefont {Zhang}}, \bibinfo {author} {\bibfnamefont {B.}~\bibnamefont {Zhou}}, \bibinfo {author} {\bibfnamefont {Y.}~\bibnamefont {Zhang}}, \bibinfo {author} {\bibfnamefont {Y.~F.}\ \bibnamefont {Guo}}, \bibinfo {author} {\bibfnamefont {M.}~\bibnamefont {Rahn}}, \bibinfo {author} {\bibfnamefont {D.}~\bibnamefont {Prabhakaran}}, \bibinfo {author} {\bibfnamefont {Z.}~\bibnamefont {Hussain}}, \bibinfo {author} {\bibfnamefont {S.~K.}\ \bibnamefont {Mo}}, \bibinfo {author} {\bibfnamefont {C.}~\bibnamefont {Felser}}, \bibinfo {author} {\bibfnamefont {B.}~\bibnamefont {Yan}},\ and\ \bibinfo {author} {\bibfnamefont {Y.~L.}\ \bibnamefont {Chen}},\ }\href@noop {} {\bibfield  {journal} {\bibinfo  {journal} {Nat. Phys.}\ }\textbf {\bibinfo {volume} {11}},\ \bibinfo {pages} {728} (\bibinfo {year} {2015})}\BibitemShut {NoStop}%
\bibitem [{\citenamefont {Liu}\ \emph {et~al.}(2016)\citenamefont {Liu}, \citenamefont {Yang}, \citenamefont {Sun}, \citenamefont {Zhang}, \citenamefont {Peng}, \citenamefont {Yang}, \citenamefont {Chen}, \citenamefont {Zhang}, \citenamefont {Guo}, \citenamefont {Prabhakaran}, \citenamefont {Schmidt}, \citenamefont {Hussain}, \citenamefont {Mo}, \citenamefont {Felser}, \citenamefont {Yan},\ and\ \citenamefont {Chen}}]{liu16}%
  \BibitemOpen
  \bibfield  {author} {\bibinfo {author} {\bibfnamefont {Z.~K.}\ \bibnamefont {Liu}}, \bibinfo {author} {\bibfnamefont {L.~X.}\ \bibnamefont {Yang}}, \bibinfo {author} {\bibfnamefont {Y.}~\bibnamefont {Sun}}, \bibinfo {author} {\bibfnamefont {T.}~\bibnamefont {Zhang}}, \bibinfo {author} {\bibfnamefont {H.}~\bibnamefont {Peng}}, \bibinfo {author} {\bibfnamefont {H.~F.}\ \bibnamefont {Yang}}, \bibinfo {author} {\bibfnamefont {C.}~\bibnamefont {Chen}}, \bibinfo {author} {\bibfnamefont {Y.}~\bibnamefont {Zhang}}, \bibinfo {author} {\bibfnamefont {Y.~F.}\ \bibnamefont {Guo}}, \bibinfo {author} {\bibfnamefont {D.}~\bibnamefont {Prabhakaran}}, \bibinfo {author} {\bibfnamefont {M.}~\bibnamefont {Schmidt}}, \bibinfo {author} {\bibfnamefont {Z.}~\bibnamefont {Hussain}}, \bibinfo {author} {\bibfnamefont {S.~K.}\ \bibnamefont {Mo}}, \bibinfo {author} {\bibfnamefont {C.}~\bibnamefont {Felser}}, \bibinfo {author} {\bibfnamefont {B.}~\bibnamefont {Yan}},\ and\ \bibinfo {author} {\bibfnamefont {Y.~L.}\ \bibnamefont {Chen}},\ }\href@noop {} {\bibfield  {journal} {\bibinfo  {journal} {Nat. Mater.}\ }\textbf {\bibinfo {volume} {15}},\ \bibinfo {pages} {27} (\bibinfo {year} {2016})}\BibitemShut {NoStop}%
\bibitem [{\citenamefont {Fang}\ \emph {et~al.}(2016)\citenamefont {Fang}, \citenamefont {Weng}, \citenamefont {Dai},\ and\ \citenamefont {Fang}}]{fang16}%
  \BibitemOpen
  \bibfield  {author} {\bibinfo {author} {\bibfnamefont {C.}~\bibnamefont {Fang}}, \bibinfo {author} {\bibfnamefont {H.}~\bibnamefont {Weng}}, \bibinfo {author} {\bibfnamefont {X.}~\bibnamefont {Dai}},\ and\ \bibinfo {author} {\bibfnamefont {Z.}~\bibnamefont {Fang}},\ }\href@noop {} {\bibfield  {journal} {\bibinfo  {journal} {Chin. Phys. B}\ }\textbf {\bibinfo {volume} {25}},\ \bibinfo {pages} {117106} (\bibinfo {year} {2016})}\BibitemShut {NoStop}%
\bibitem [{\citenamefont {Yu}\ \emph {et~al.}(2017)\citenamefont {Yu}, \citenamefont {Fang}, \citenamefont {Dai},\ and\ \citenamefont {Weng}}]{yu17}%
  \BibitemOpen
  \bibfield  {author} {\bibinfo {author} {\bibfnamefont {R.}~\bibnamefont {Yu}}, \bibinfo {author} {\bibfnamefont {Z.}~\bibnamefont {Fang}}, \bibinfo {author} {\bibfnamefont {X.}~\bibnamefont {Dai}},\ and\ \bibinfo {author} {\bibfnamefont {H.}~\bibnamefont {Weng}},\ }\href@noop {} {\bibfield  {journal} {\bibinfo  {journal} {Front. Phys.}\ }\textbf {\bibinfo {volume} {12}},\ \bibinfo {pages} {127202} (\bibinfo {year} {2017})}\BibitemShut {NoStop}%
\bibitem [{\citenamefont {Schoop}\ \emph {et~al.}(2016)\citenamefont {Schoop}, \citenamefont {Ali}, \citenamefont {Strasser}, \citenamefont {Topp}, \citenamefont {Varykhalov}, \citenamefont {Marchenko}, \citenamefont {Duppel}, \citenamefont {Parkin}, \citenamefont {Lotsch},\ and\ \citenamefont {Ast}}]{schoop16}%
  \BibitemOpen
  \bibfield  {author} {\bibinfo {author} {\bibfnamefont {L.~M.}\ \bibnamefont {Schoop}}, \bibinfo {author} {\bibfnamefont {M.~N.}\ \bibnamefont {Ali}}, \bibinfo {author} {\bibfnamefont {C.}~\bibnamefont {Strasser}}, \bibinfo {author} {\bibfnamefont {A.}~\bibnamefont {Topp}}, \bibinfo {author} {\bibfnamefont {A.}~\bibnamefont {Varykhalov}}, \bibinfo {author} {\bibfnamefont {D.}~\bibnamefont {Marchenko}}, \bibinfo {author} {\bibfnamefont {V.}~\bibnamefont {Duppel}}, \bibinfo {author} {\bibfnamefont {S.~S.}\ \bibnamefont {Parkin}}, \bibinfo {author} {\bibfnamefont {B.~V.}\ \bibnamefont {Lotsch}},\ and\ \bibinfo {author} {\bibfnamefont {C.~R.}\ \bibnamefont {Ast}},\ }\href@noop {} {\bibfield  {journal} {\bibinfo  {journal} {Nat. Commun.}\ }\textbf {\bibinfo {volume} {7}},\ \bibinfo {pages} {11696} (\bibinfo {year} {2016})}\BibitemShut {NoStop}%
\bibitem [{\citenamefont {Fang}\ \emph {et~al.}(2015)\citenamefont {Fang}, \citenamefont {Chen}, \citenamefont {Kee},\ and\ \citenamefont {Fu}}]{fang15}%
  \BibitemOpen
  \bibfield  {author} {\bibinfo {author} {\bibfnamefont {C.}~\bibnamefont {Fang}}, \bibinfo {author} {\bibfnamefont {Y.}~\bibnamefont {Chen}}, \bibinfo {author} {\bibfnamefont {H.-Y.}\ \bibnamefont {Kee}},\ and\ \bibinfo {author} {\bibfnamefont {L.}~\bibnamefont {Fu}},\ }\href@noop {} {\bibfield  {journal} {\bibinfo  {journal} {Phys. Rev. B}\ }\textbf {\bibinfo {volume} {92}},\ \bibinfo {pages} {081201(R)} (\bibinfo {year} {2015})}\BibitemShut {NoStop}%
\bibitem [{\citenamefont {Bzdusek}\ \emph {et~al.}(2016)\citenamefont {Bzdusek}, \citenamefont {Wu}, \citenamefont {Ruegg}, \citenamefont {Sigrist},\ and\ \citenamefont {Soluyanov}}]{bzdusek16}%
  \BibitemOpen
  \bibfield  {author} {\bibinfo {author} {\bibfnamefont {T.}~\bibnamefont {Bzdusek}}, \bibinfo {author} {\bibfnamefont {Q.}~\bibnamefont {Wu}}, \bibinfo {author} {\bibfnamefont {A.}~\bibnamefont {Ruegg}}, \bibinfo {author} {\bibfnamefont {M.}~\bibnamefont {Sigrist}},\ and\ \bibinfo {author} {\bibfnamefont {A.~A.}\ \bibnamefont {Soluyanov}},\ }\href@noop {} {\bibfield  {journal} {\bibinfo  {journal} {Nature (London)}\ }\textbf {\bibinfo {volume} {538}},\ \bibinfo {pages} {75} (\bibinfo {year} {2016})}\BibitemShut {NoStop}%
\bibitem [{\citenamefont {Yang}\ \emph {et~al.}(2018)\citenamefont {Yang}, \citenamefont {Yang}, \citenamefont {Derunova}, \citenamefont {Parkin}, \citenamefont {Yan},\ and\ \citenamefont {Ali}}]{yang18}%
  \BibitemOpen
  \bibfield  {author} {\bibinfo {author} {\bibfnamefont {S.-Y.}\ \bibnamefont {Yang}}, \bibinfo {author} {\bibfnamefont {H.}~\bibnamefont {Yang}}, \bibinfo {author} {\bibfnamefont {E.}~\bibnamefont {Derunova}}, \bibinfo {author} {\bibfnamefont {S.~S.~P.}\ \bibnamefont {Parkin}}, \bibinfo {author} {\bibfnamefont {B.}~\bibnamefont {Yan}},\ and\ \bibinfo {author} {\bibfnamefont {M.~N.}\ \bibnamefont {Ali}},\ }\href@noop {} {\bibfield  {journal} {\bibinfo  {journal} {Adv. Phys.: X}\ }\textbf {\bibinfo {volume} {3}},\ \bibinfo {pages} {1414631} (\bibinfo {year} {2018})}\BibitemShut {NoStop}%
\bibitem [{\citenamefont {Hu}\ \emph {et~al.}(2016)\citenamefont {Hu}, \citenamefont {Tang}, \citenamefont {Liu}, \citenamefont {Liu}, \citenamefont {Zhu}, \citenamefont {Graf}, \citenamefont {Myhro}, \citenamefont {Tran}, \citenamefont {Lau}, \citenamefont {Wei},\ and\ \citenamefont {Mao}}]{hu16}%
  \BibitemOpen
  \bibfield  {author} {\bibinfo {author} {\bibfnamefont {J.}~\bibnamefont {Hu}}, \bibinfo {author} {\bibfnamefont {Z.}~\bibnamefont {Tang}}, \bibinfo {author} {\bibfnamefont {J.}~\bibnamefont {Liu}}, \bibinfo {author} {\bibfnamefont {X.}~\bibnamefont {Liu}}, \bibinfo {author} {\bibfnamefont {Y.}~\bibnamefont {Zhu}}, \bibinfo {author} {\bibfnamefont {D.}~\bibnamefont {Graf}}, \bibinfo {author} {\bibfnamefont {K.}~\bibnamefont {Myhro}}, \bibinfo {author} {\bibfnamefont {S.}~\bibnamefont {Tran}}, \bibinfo {author} {\bibfnamefont {C.~N.}\ \bibnamefont {Lau}}, \bibinfo {author} {\bibfnamefont {J.}~\bibnamefont {Wei}},\ and\ \bibinfo {author} {\bibfnamefont {Z.}~\bibnamefont {Mao}},\ }\href@noop {} {\bibfield  {journal} {\bibinfo  {journal} {Phys. Rev. Lett.}\ }\textbf {\bibinfo {volume} {117}},\ \bibinfo {pages} {016602} (\bibinfo {year} {2016})}\BibitemShut {NoStop}%
\bibitem [{\citenamefont {Hao}\ \emph {et~al.}(2021)\citenamefont {Hao}, \citenamefont {Chen}, \citenamefont {Wang}, \citenamefont {Li}, \citenamefont {Ma}, \citenamefont {Hao}, \citenamefont {Lu}, \citenamefont {Shen}, \citenamefont {Jiang}, \citenamefont {Liu} \emph {et~al.}}]{hao2021multiple}%
  \BibitemOpen
  \bibfield  {author} {\bibinfo {author} {\bibfnamefont {Z.}~\bibnamefont {Hao}}, \bibinfo {author} {\bibfnamefont {W.}~\bibnamefont {Chen}}, \bibinfo {author} {\bibfnamefont {Y.}~\bibnamefont {Wang}}, \bibinfo {author} {\bibfnamefont {J.}~\bibnamefont {Li}}, \bibinfo {author} {\bibfnamefont {X.-M.}\ \bibnamefont {Ma}}, \bibinfo {author} {\bibfnamefont {Y.-J.}\ \bibnamefont {Hao}}, \bibinfo {author} {\bibfnamefont {R.}~\bibnamefont {Lu}}, \bibinfo {author} {\bibfnamefont {Z.}~\bibnamefont {Shen}}, \bibinfo {author} {\bibfnamefont {Z.}~\bibnamefont {Jiang}}, \bibinfo {author} {\bibfnamefont {W.}~\bibnamefont {Liu}}, \emph {et~al.},\ }\href@noop {} {\bibfield  {journal} {\bibinfo  {journal} {Physical Review B}\ }\textbf {\bibinfo {volume} {104}},\ \bibinfo {pages} {115158} (\bibinfo {year} {2021})}\BibitemShut {NoStop}%
\bibitem [{\citenamefont {Xu}\ \emph {et~al.}(2020)\citenamefont {Xu}, \citenamefont {Liu}, \citenamefont {Cai}, \citenamefont {Li}, \citenamefont {Jiao}, \citenamefont {Li}, \citenamefont {Zhang}, \citenamefont {Zhou}, \citenamefont {Qian}, \citenamefont {Jiang} \emph {et~al.}}]{xu2020anisotropic}%
  \BibitemOpen
  \bibfield  {author} {\bibinfo {author} {\bibfnamefont {C.}~\bibnamefont {Xu}}, \bibinfo {author} {\bibfnamefont {Y.}~\bibnamefont {Liu}}, \bibinfo {author} {\bibfnamefont {P.}~\bibnamefont {Cai}}, \bibinfo {author} {\bibfnamefont {B.}~\bibnamefont {Li}}, \bibinfo {author} {\bibfnamefont {W.}~\bibnamefont {Jiao}}, \bibinfo {author} {\bibfnamefont {Y.}~\bibnamefont {Li}}, \bibinfo {author} {\bibfnamefont {J.}~\bibnamefont {Zhang}}, \bibinfo {author} {\bibfnamefont {W.}~\bibnamefont {Zhou}}, \bibinfo {author} {\bibfnamefont {B.}~\bibnamefont {Qian}}, \bibinfo {author} {\bibfnamefont {X.}~\bibnamefont {Jiang}}, \emph {et~al.},\ }\href@noop {} {\bibfield  {journal} {\bibinfo  {journal} {The Journal of Physical Chemistry Letters}\ }\textbf {\bibinfo {volume} {11}},\ \bibinfo {pages} {7782} (\bibinfo {year} {2020})}\BibitemShut {NoStop}%
\bibitem [{\citenamefont {Chen}\ \emph {et~al.}(2021)\citenamefont {Chen}, \citenamefont {Wu}, \citenamefont {Zhang}, \citenamefont {Zhang}, \citenamefont {Nie}, \citenamefont {Qin}, \citenamefont {Han}, \citenamefont {Xi}, \citenamefont {Ma}, \citenamefont {Kan}, \citenamefont {Zhou}, \citenamefont {Yang}, \citenamefont {Zhu}, \citenamefont {Ning},\ and\ \citenamefont {Tian}}]{chen21}%
  \BibitemOpen
  \bibfield  {author} {\bibinfo {author} {\bibfnamefont {Z.}~\bibnamefont {Chen}}, \bibinfo {author} {\bibfnamefont {M.}~\bibnamefont {Wu}}, \bibinfo {author} {\bibfnamefont {Y.}~\bibnamefont {Zhang}}, \bibinfo {author} {\bibfnamefont {J.}~\bibnamefont {Zhang}}, \bibinfo {author} {\bibfnamefont {Y.}~\bibnamefont {Nie}}, \bibinfo {author} {\bibfnamefont {Y.}~\bibnamefont {Qin}}, \bibinfo {author} {\bibfnamefont {Y.}~\bibnamefont {Han}}, \bibinfo {author} {\bibfnamefont {C.}~\bibnamefont {Xi}}, \bibinfo {author} {\bibfnamefont {S.}~\bibnamefont {Ma}}, \bibinfo {author} {\bibfnamefont {X.}~\bibnamefont {Kan}}, \bibinfo {author} {\bibfnamefont {J.}~\bibnamefont {Zhou}}, \bibinfo {author} {\bibfnamefont {X.}~\bibnamefont {Yang}}, \bibinfo {author} {\bibfnamefont {X.}~\bibnamefont {Zhu}}, \bibinfo {author} {\bibfnamefont {W.}~\bibnamefont {Ning}},\ and\ \bibinfo {author} {\bibfnamefont {M.}~\bibnamefont {Tian}},\ }\href@noop {} {\bibfield  {journal} {\bibinfo  {journal} {Phys. Rev. B}\ }\textbf {\bibinfo {volume} {103}},\ \bibinfo {pages} {035105} (\bibinfo {year} {2021})}\BibitemShut {NoStop}%
\bibitem [{\citenamefont {Ye}\ \emph {et~al.}(2022)\citenamefont {Ye}, \citenamefont {Gao}, \citenamefont {Li}, \citenamefont {Liang},\ and\ \citenamefont {Cao}}]{ye22}%
  \BibitemOpen
  \bibfield  {author} {\bibinfo {author} {\bibfnamefont {R.}~\bibnamefont {Ye}}, \bibinfo {author} {\bibfnamefont {T.}~\bibnamefont {Gao}}, \bibinfo {author} {\bibfnamefont {H.}~\bibnamefont {Li}}, \bibinfo {author} {\bibfnamefont {X.}~\bibnamefont {Liang}},\ and\ \bibinfo {author} {\bibfnamefont {G.}~\bibnamefont {Cao}},\ }\href@noop {} {\bibfield  {journal} {\bibinfo  {journal} {AIP Advances}\ }\textbf {\bibinfo {volume} {12}},\ \bibinfo {pages} {045104} (\bibinfo {year} {2022})}\BibitemShut {NoStop}%
\bibitem [{\citenamefont {Liimatta}\ and\ \citenamefont {Ibers}(1989)}]{liimatta89}%
  \BibitemOpen
  \bibfield  {author} {\bibinfo {author} {\bibfnamefont {C.~E.~W.}\ \bibnamefont {Liimatta}}\ and\ \bibinfo {author} {\bibfnamefont {J.~A.}\ \bibnamefont {Ibers}},\ }\href@noop {} {\bibfield  {journal} {\bibinfo  {journal} {J. Solid State Chem.}\ }\textbf {\bibinfo {volume} {78}},\ \bibinfo {pages} {7} (\bibinfo {year} {1989})}\BibitemShut {NoStop}%
\bibitem [{\citenamefont {Blaha}\ \emph {et~al.}(2001)\citenamefont {Blaha}, \citenamefont {Schwarz}, \citenamefont {Madsen}, \citenamefont {Kvasnicka}, \citenamefont {Luitz} \emph {et~al.}}]{blaha2001wien2k}%
  \BibitemOpen
  \bibfield  {author} {\bibinfo {author} {\bibfnamefont {P.}~\bibnamefont {Blaha}}, \bibinfo {author} {\bibfnamefont {K.}~\bibnamefont {Schwarz}}, \bibinfo {author} {\bibfnamefont {G.~K.}\ \bibnamefont {Madsen}}, \bibinfo {author} {\bibfnamefont {D.}~\bibnamefont {Kvasnicka}}, \bibinfo {author} {\bibfnamefont {J.}~\bibnamefont {Luitz}}, \emph {et~al.},\ }\href@noop {} {\bibfield  {journal} {\bibinfo  {journal} {An augmented plane wave+ local orbitals program for calculating crystal properties}\ }\textbf {\bibinfo {volume} {60}} (\bibinfo {year} {2001})}\BibitemShut {NoStop}%
\bibitem [{\citenamefont {Perdew}\ \emph {et~al.}(1996)\citenamefont {Perdew}, \citenamefont {Burke},\ and\ \citenamefont {Ernzerhof}}]{Perdew96}%
  \BibitemOpen
  \bibfield  {author} {\bibinfo {author} {\bibfnamefont {J.~P.}\ \bibnamefont {Perdew}}, \bibinfo {author} {\bibfnamefont {K.}~\bibnamefont {Burke}},\ and\ \bibinfo {author} {\bibfnamefont {M.}~\bibnamefont {Ernzerhof}},\ }\href@noop {} {\bibfield  {journal} {\bibinfo  {journal} {Phys. Rev. Lett.}\ }\textbf {\bibinfo {volume} {77}},\ \bibinfo {pages} {3865} (\bibinfo {year} {1996})}\BibitemShut {NoStop}%
\bibitem [{\citenamefont {Young}\ \emph {et~al.}(2012)\citenamefont {Young}, \citenamefont {Zaheer}, \citenamefont {Teo}, \citenamefont {Kane}, \citenamefont {Mele},\ and\ \citenamefont {Rappe}}]{young2012dirac}%
  \BibitemOpen
  \bibfield  {author} {\bibinfo {author} {\bibfnamefont {S.~M.}\ \bibnamefont {Young}}, \bibinfo {author} {\bibfnamefont {S.}~\bibnamefont {Zaheer}}, \bibinfo {author} {\bibfnamefont {J.~C.}\ \bibnamefont {Teo}}, \bibinfo {author} {\bibfnamefont {C.~L.}\ \bibnamefont {Kane}}, \bibinfo {author} {\bibfnamefont {E.~J.}\ \bibnamefont {Mele}},\ and\ \bibinfo {author} {\bibfnamefont {A.~M.}\ \bibnamefont {Rappe}},\ }\href@noop {} {\bibfield  {journal} {\bibinfo  {journal} {Physical review letters}\ }\textbf {\bibinfo {volume} {108}},\ \bibinfo {pages} {140405} (\bibinfo {year} {2012})}\BibitemShut {NoStop}%
\bibitem [{\citenamefont {Xiao}\ \emph {et~al.}(2022)\citenamefont {Xiao}, \citenamefont {Jiao}, \citenamefont {Lin}, \citenamefont {Jiang}, \citenamefont {Yang}, \citenamefont {He}, \citenamefont {Jiang}, \citenamefont {Yang}, \citenamefont {Liu}, \citenamefont {Ye} \emph {et~al.}}]{xiao2022dirac}%
  \BibitemOpen
  \bibfield  {author} {\bibinfo {author} {\bibfnamefont {S.}~\bibnamefont {Xiao}}, \bibinfo {author} {\bibfnamefont {W.-H.}\ \bibnamefont {Jiao}}, \bibinfo {author} {\bibfnamefont {Y.}~\bibnamefont {Lin}}, \bibinfo {author} {\bibfnamefont {Q.}~\bibnamefont {Jiang}}, \bibinfo {author} {\bibfnamefont {X.}~\bibnamefont {Yang}}, \bibinfo {author} {\bibfnamefont {Y.}~\bibnamefont {He}}, \bibinfo {author} {\bibfnamefont {Z.}~\bibnamefont {Jiang}}, \bibinfo {author} {\bibfnamefont {Y.}~\bibnamefont {Yang}}, \bibinfo {author} {\bibfnamefont {Z.}~\bibnamefont {Liu}}, \bibinfo {author} {\bibfnamefont {M.}~\bibnamefont {Ye}}, \emph {et~al.},\ }\href@noop {} {\bibfield  {journal} {\bibinfo  {journal} {Physical Review B}\ }\textbf {\bibinfo {volume} {105}},\ \bibinfo {pages} {195145} (\bibinfo {year} {2022})}\BibitemShut {NoStop}%
\bibitem [{\citenamefont {Badger}\ \emph {et~al.}(2022)\citenamefont {Badger}, \citenamefont {Quan}, \citenamefont {Staab}, \citenamefont {Sumita}, \citenamefont {Rossi}, \citenamefont {Devlin}, \citenamefont {Neubauer}, \citenamefont {Shulman}, \citenamefont {Fettinger}, \citenamefont {Klavins} \emph {et~al.}}]{badger2022dirac}%
  \BibitemOpen
  \bibfield  {author} {\bibinfo {author} {\bibfnamefont {J.~R.}\ \bibnamefont {Badger}}, \bibinfo {author} {\bibfnamefont {Y.}~\bibnamefont {Quan}}, \bibinfo {author} {\bibfnamefont {M.~C.}\ \bibnamefont {Staab}}, \bibinfo {author} {\bibfnamefont {S.}~\bibnamefont {Sumita}}, \bibinfo {author} {\bibfnamefont {A.}~\bibnamefont {Rossi}}, \bibinfo {author} {\bibfnamefont {K.~P.}\ \bibnamefont {Devlin}}, \bibinfo {author} {\bibfnamefont {K.}~\bibnamefont {Neubauer}}, \bibinfo {author} {\bibfnamefont {D.~S.}\ \bibnamefont {Shulman}}, \bibinfo {author} {\bibfnamefont {J.~C.}\ \bibnamefont {Fettinger}}, \bibinfo {author} {\bibfnamefont {P.}~\bibnamefont {Klavins}}, \emph {et~al.},\ }\href@noop {} {\bibfield  {journal} {\bibinfo  {journal} {Communications Physics}\ }\textbf {\bibinfo {volume} {5}},\ \bibinfo {pages} {22} (\bibinfo {year} {2022})}\BibitemShut {NoStop}%
\bibitem [{\citenamefont {Wang}\ \emph {et~al.}(2022)\citenamefont {Wang}, \citenamefont {Wang}, \citenamefont {Yang}, \citenamefont {Fan}, \citenamefont {Zheng}, \citenamefont {Wang},\ and\ \citenamefont {Wu}}]{wang2022symmetry}%
  \BibitemOpen
  \bibfield  {author} {\bibinfo {author} {\bibfnamefont {M.}~\bibnamefont {Wang}}, \bibinfo {author} {\bibfnamefont {Y.}~\bibnamefont {Wang}}, \bibinfo {author} {\bibfnamefont {Z.}~\bibnamefont {Yang}}, \bibinfo {author} {\bibfnamefont {J.}~\bibnamefont {Fan}}, \bibinfo {author} {\bibfnamefont {B.}~\bibnamefont {Zheng}}, \bibinfo {author} {\bibfnamefont {R.}~\bibnamefont {Wang}},\ and\ \bibinfo {author} {\bibfnamefont {X.}~\bibnamefont {Wu}},\ }\href@noop {} {\bibfield  {journal} {\bibinfo  {journal} {Physical Review B}\ }\textbf {\bibinfo {volume} {105}},\ \bibinfo {pages} {174309} (\bibinfo {year} {2022})}\BibitemShut {NoStop}%
\bibitem [{\citenamefont {Jiao}\ \emph {et~al.}(2020)\citenamefont {Jiao}, \citenamefont {Xie}, \citenamefont {Liu}, \citenamefont {Xu}, \citenamefont {Li}, \citenamefont {Xu}, \citenamefont {Liu}, \citenamefont {Zhou}, \citenamefont {Li}, \citenamefont {Yang}, \citenamefont {Jiang}, \citenamefont {Luo}, \citenamefont {Zhu},\ and\ \citenamefont {Cao}}]{jiao20}%
  \BibitemOpen
  \bibfield  {author} {\bibinfo {author} {\bibfnamefont {W.-H.}\ \bibnamefont {Jiao}}, \bibinfo {author} {\bibfnamefont {X.-M.}\ \bibnamefont {Xie}}, \bibinfo {author} {\bibfnamefont {Y.}~\bibnamefont {Liu}}, \bibinfo {author} {\bibfnamefont {X.}~\bibnamefont {Xu}}, \bibinfo {author} {\bibfnamefont {B.}~\bibnamefont {Li}}, \bibinfo {author} {\bibfnamefont {C.-Q.}\ \bibnamefont {Xu}}, \bibinfo {author} {\bibfnamefont {J.-Y.}\ \bibnamefont {Liu}}, \bibinfo {author} {\bibfnamefont {W.}~\bibnamefont {Zhou}}, \bibinfo {author} {\bibfnamefont {Y.-K.}\ \bibnamefont {Li}}, \bibinfo {author} {\bibfnamefont {H.-Y.}\ \bibnamefont {Yang}}, \bibinfo {author} {\bibfnamefont {S.}~\bibnamefont {Jiang}}, \bibinfo {author} {\bibfnamefont {Y.}~\bibnamefont {Luo}}, \bibinfo {author} {\bibfnamefont {Z.-W.}\ \bibnamefont {Zhu}},\ and\ \bibinfo {author} {\bibfnamefont {G.-H.}\ \bibnamefont {Cao}},\ }\href@noop {} {\bibfield  {journal} {\bibinfo  {journal} {Phys. Rev. B}\ }\textbf {\bibinfo {volume} {102}},\ \bibinfo {pages} {075141} (\bibinfo {year} {2020})}\BibitemShut {NoStop}%
\bibitem [{\citenamefont {Jiao}\ \emph {et~al.}(2021)\citenamefont {Jiao}, \citenamefont {Xiao}, \citenamefont {Li}, \citenamefont {Xu}, \citenamefont {Xie}, \citenamefont {Qiu}, \citenamefont {Xu}, \citenamefont {Liu}, \citenamefont {Song}, \citenamefont {Zhou}, \citenamefont {Zhai}, \citenamefont {Ke}, \citenamefont {He},\ and\ \citenamefont {Cao}}]{jiao21}%
  \BibitemOpen
  \bibfield  {author} {\bibinfo {author} {\bibfnamefont {W.-H.}\ \bibnamefont {Jiao}}, \bibinfo {author} {\bibfnamefont {S.}~\bibnamefont {Xiao}}, \bibinfo {author} {\bibfnamefont {B.}~\bibnamefont {Li}}, \bibinfo {author} {\bibfnamefont {C.}~\bibnamefont {Xu}}, \bibinfo {author} {\bibfnamefont {X.-M.}\ \bibnamefont {Xie}}, \bibinfo {author} {\bibfnamefont {H.-Q.}\ \bibnamefont {Qiu}}, \bibinfo {author} {\bibfnamefont {X.}~\bibnamefont {Xu}}, \bibinfo {author} {\bibfnamefont {Y.}~\bibnamefont {Liu}}, \bibinfo {author} {\bibfnamefont {S.-J.}\ \bibnamefont {Song}}, \bibinfo {author} {\bibfnamefont {W.}~\bibnamefont {Zhou}}, \bibinfo {author} {\bibfnamefont {H.-F.}\ \bibnamefont {Zhai}}, \bibinfo {author} {\bibfnamefont {X.}~\bibnamefont {Ke}}, \bibinfo {author} {\bibfnamefont {S.}~\bibnamefont {He}},\ and\ \bibinfo {author} {\bibfnamefont {G.-H.}\ \bibnamefont {Cao}},\ }\href@noop {} {\bibfield  {journal} {\bibinfo  {journal} {Phys. Rev. B}\ }\textbf {\bibinfo {volume} {103}},\ \bibinfo {pages} {125150} (\bibinfo {year} {2021})}\BibitemShut {NoStop}%
\bibitem [{\citenamefont {Shoenberg}(1984)}]{shoenberg84}%
  \BibitemOpen
  \bibfield  {author} {\bibinfo {author} {\bibfnamefont {D.}~\bibnamefont {Shoenberg}},\ }\href@noop {} {\emph {\bibinfo {title} {Magnetic oscillations in metals}}}\ (\bibinfo  {publisher} {Cambridge university press},\ \bibinfo {year} {1984})\BibitemShut {NoStop}%
\bibitem [{\citenamefont {Rourke}\ and\ \citenamefont {Julian}(2012)}]{julian2012numerical}%
  \BibitemOpen
  \bibfield  {author} {\bibinfo {author} {\bibfnamefont {P.}~\bibnamefont {Rourke}}\ and\ \bibinfo {author} {\bibfnamefont {S.}~\bibnamefont {Julian}},\ }\href@noop {} {\bibfield  {journal} {\bibinfo  {journal} {Computer Physics Communications}\ }\textbf {\bibinfo {volume} {183}},\ \bibinfo {pages} {324} (\bibinfo {year} {2012})}\BibitemShut {NoStop}%
\bibitem [{\citenamefont {Hu}\ \emph {et~al.}(2017)\citenamefont {Hu}, \citenamefont {Tang}, \citenamefont {Liu}, \citenamefont {Zhu}, \citenamefont {Wei},\ and\ \citenamefont {Mao}}]{hu2017nearly}%
  \BibitemOpen
  \bibfield  {author} {\bibinfo {author} {\bibfnamefont {J.}~\bibnamefont {Hu}}, \bibinfo {author} {\bibfnamefont {Z.}~\bibnamefont {Tang}}, \bibinfo {author} {\bibfnamefont {J.}~\bibnamefont {Liu}}, \bibinfo {author} {\bibfnamefont {Y.}~\bibnamefont {Zhu}}, \bibinfo {author} {\bibfnamefont {J.}~\bibnamefont {Wei}},\ and\ \bibinfo {author} {\bibfnamefont {Z.}~\bibnamefont {Mao}},\ }\href@noop {} {\bibfield  {journal} {\bibinfo  {journal} {Physical Review B}\ }\textbf {\bibinfo {volume} {96}},\ \bibinfo {pages} {045127} (\bibinfo {year} {2017})}\BibitemShut {NoStop}%
\bibitem [{\citenamefont {Daschner}\ \emph {et~al.}(2025)\citenamefont {Daschner}, \citenamefont {Kokanovi{\'c}},\ and\ \citenamefont {Grosche}}]{daschner2025mechanical}%
  \BibitemOpen
  \bibfield  {author} {\bibinfo {author} {\bibfnamefont {M.}~\bibnamefont {Daschner}}, \bibinfo {author} {\bibfnamefont {I.}~\bibnamefont {Kokanovi{\'c}}},\ and\ \bibinfo {author} {\bibfnamefont {F.~M.}\ \bibnamefont {Grosche}},\ }\href@noop {} {\bibfield  {journal} {\bibinfo  {journal} {arXiv preprint arXiv:2507.02612}\ } (\bibinfo {year} {2025})}\BibitemShut {NoStop}%
\end{thebibliography}%

\clearpage

\onecolumngrid
\appendix

\section{Methods}\label{methods}

\begin{figure*}[ht]
\includegraphics[width=\linewidth]{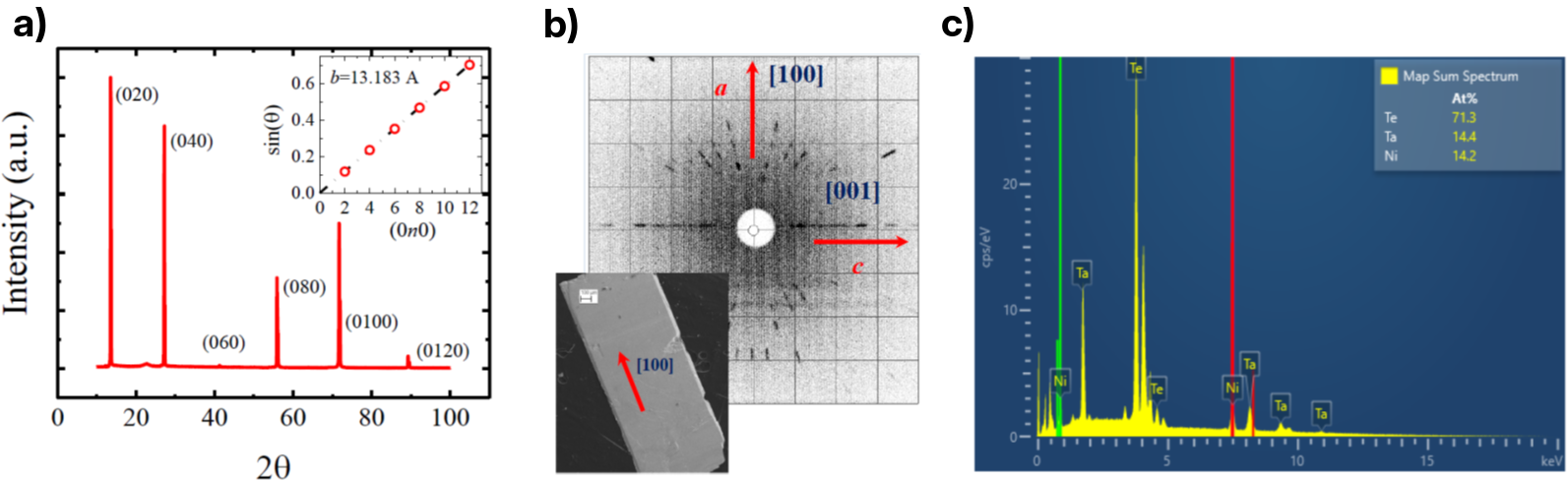}
\caption{a) Single crystal X-ray diffraction pattern. The inset shows the extracted value of the lattice parameter b. The sharp diffraction peaks indicate the high quality of our crystals. b) Laue X-ray diffraction pattern, where the inset shows an optical image of a typical crystal. c) The energy-dispersive X-ray spectrum. }
\label{SI_laue_edx}
\end{figure*}

\section{Raw Data for Magnetization Measurements}\label{raw_data_magnetization}

\begin{figure*}[ht]
  \subfloat{
        \includegraphics[width=0.98\linewidth]{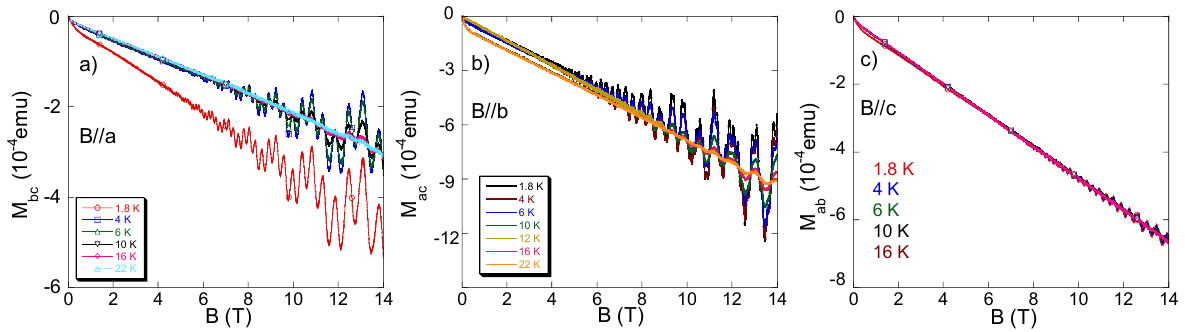}}
 	\caption{Magnetization for magnetic fields of up to \SI{14}{\tesla} at various temperatures with the magnetic field applied parallel to the crystallographic a) \emph{a}-axis,  b) \emph{b}-axis and  c) \emph{c}-axis.}
\end{figure*}

\clearpage
\section{Analysis of Magnetization Measurements}\label{analysis_magnetization}

\begin{figure}[ht]
        \includegraphics[width=0.48\columnwidth]{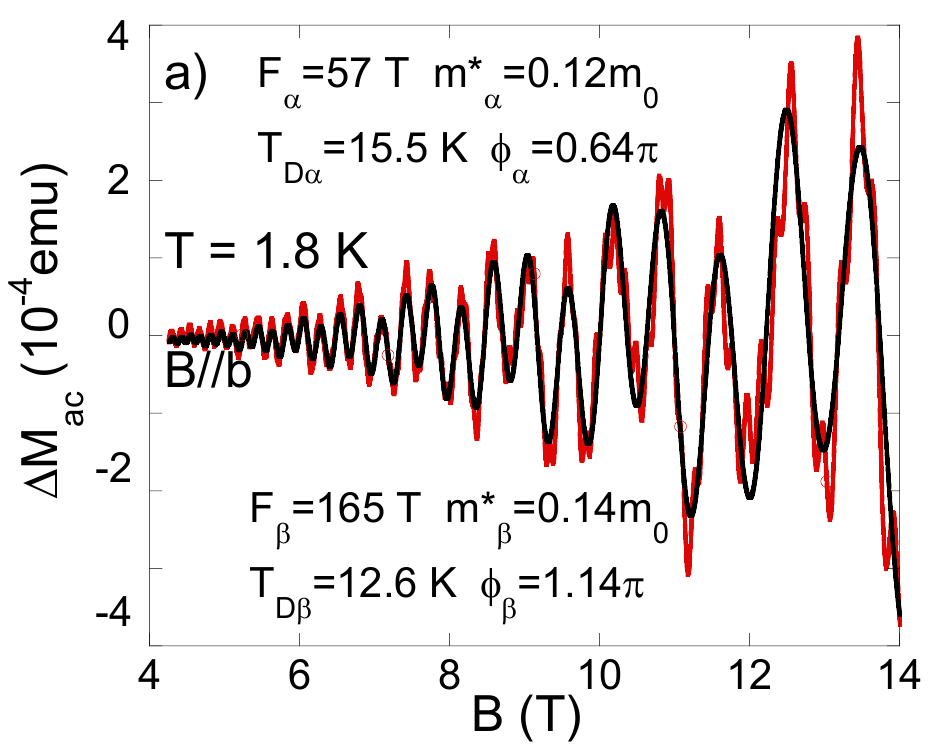}
        \includegraphics[width=0.48\columnwidth]{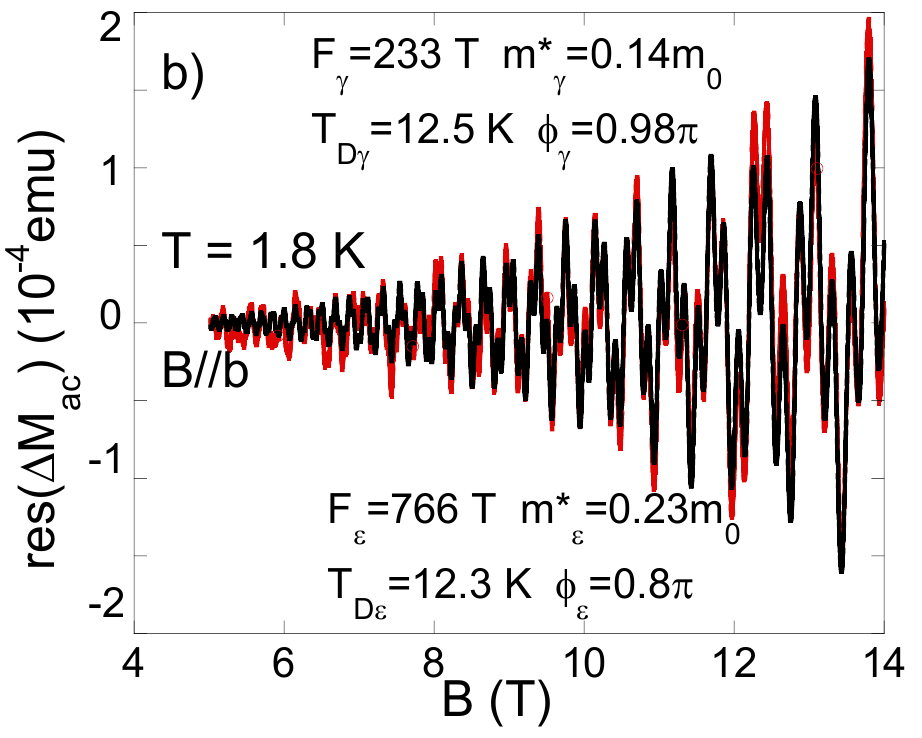}

    \caption{a) The oscillatory component of the isothermal magnetization $\Delta$M after subtracting a diamagnetic background. The magnetic field is applied parallel to the crystallographic \emph{b} axis. The black lines show the fit of the oscillation pattern assuming two bands. b) Two-band LK formula fit to the residual after subtracting the first two-band fit from the data in a). This result implies that the raw data can best be fitted with a four-band LK formula.}
    \label{SI_BIIb}
\end{figure}

\begin{figure}[ht]
        \includegraphics[width=0.48\columnwidth]{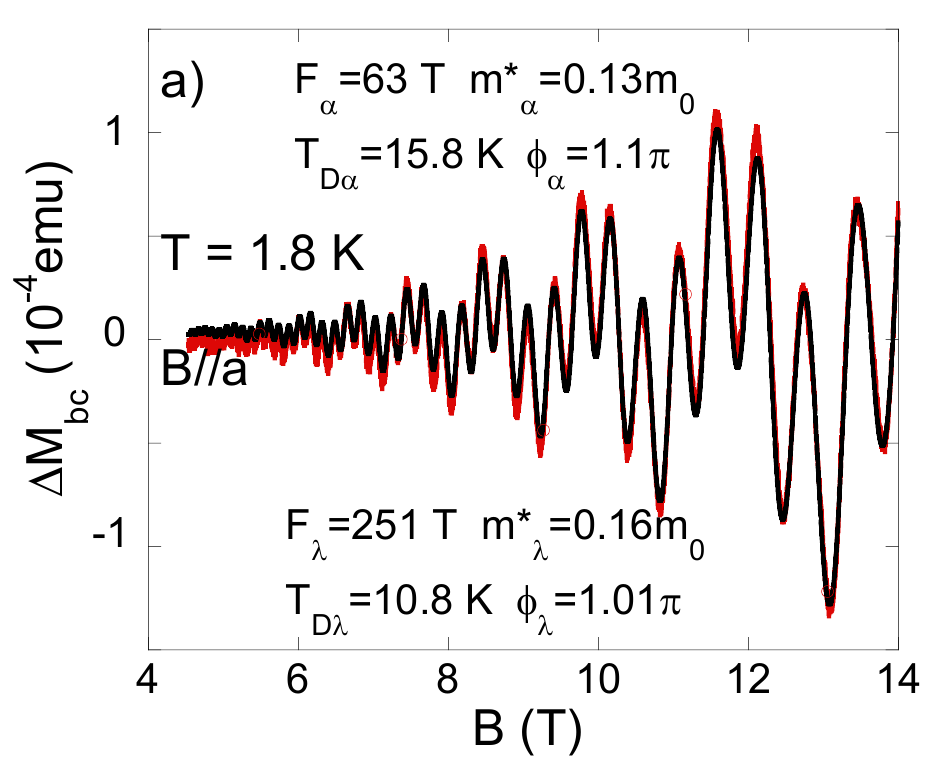}
        \includegraphics[width=0.48\columnwidth]{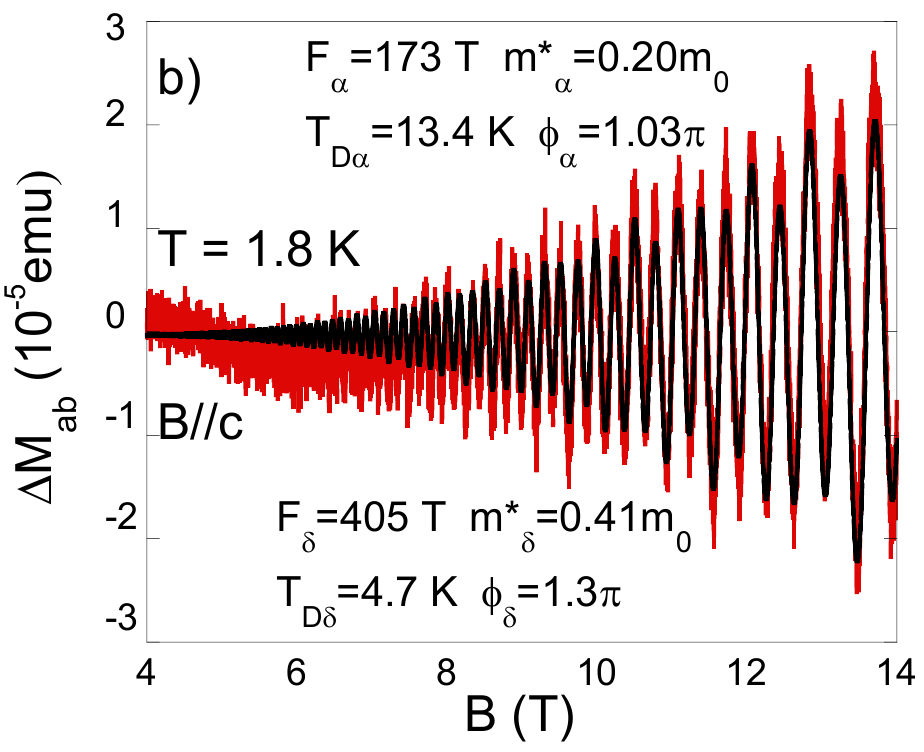}

	\caption{The oscillatory components of the isothermal magnetization $\Delta$M after subtracting the diamagnetic background. The magnetic field is applied parallel to the crystallographic a) \emph{a}, and b) \emph{c} axis. The black lines show the fit of the oscillation pattern assuming two bands. In b) another two-band fit was performed after subtracting the first two-band fit from the raw data similar to \autoref{SI_BIIb} (not shown here) to obtain values for F$_{\tau}$ and F$_{\eta}$. However, the resolution was too low to extract meaningful information about the phase shift.}
    \label{SI_BIIa-BIIc}
\end{figure}

\pagebreak

\section{Raw Data for Magnetic Torque Measurements}\label{raw_data_magnetic_torque}

\begin{figure*}[ht]
       \includegraphics[width=.45\linewidth]{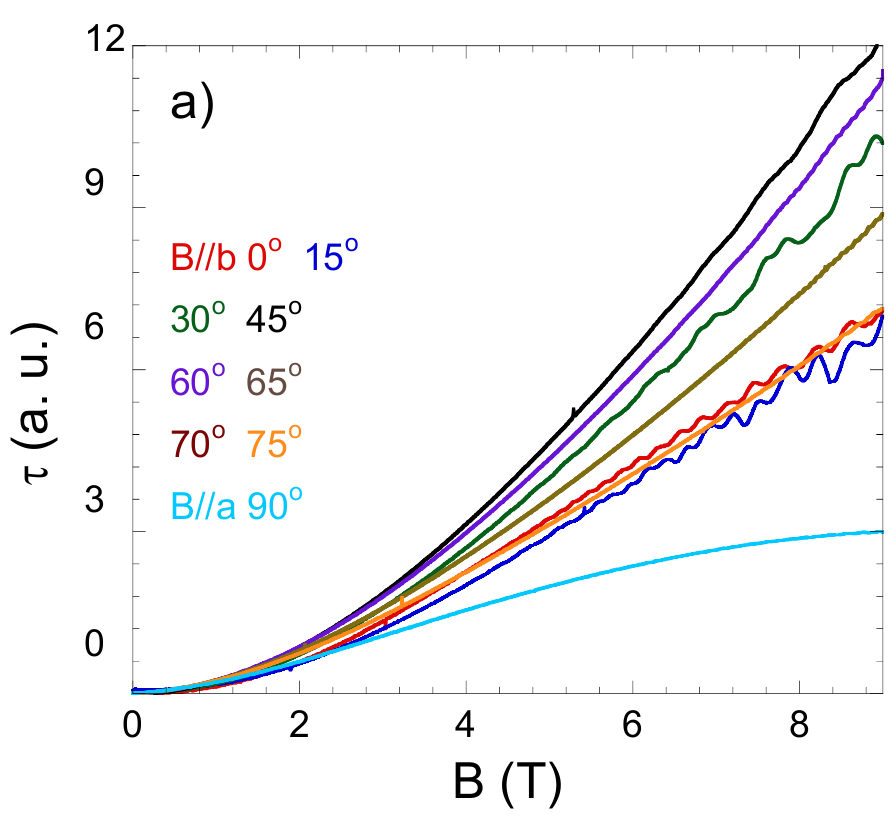}%
\hfill
       \includegraphics[width=.45\linewidth]{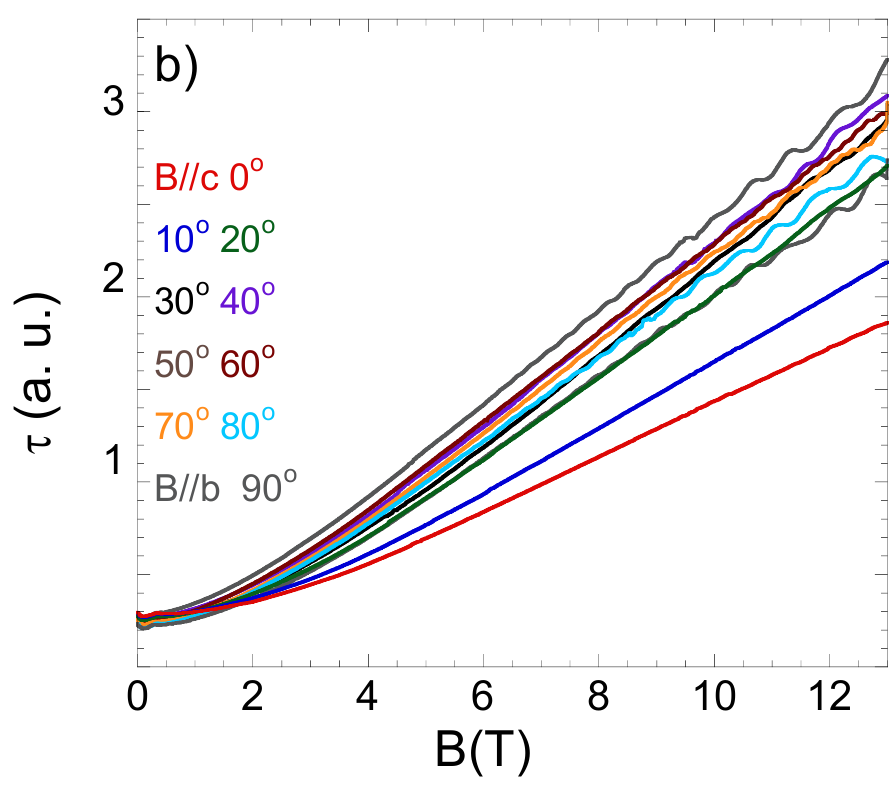}%
       \caption{a) The magnetic field sweeps of magnetic torque at various angles at \SI{1.8}{\kelvin} a) for \textbf{B}$\parallel$\emph{a}-\emph{b} plane, and b) for \textbf{B}$\parallel$\emph{b}-\emph{c} plane.  }
  \label{SI_backgroundsubtr+BIIbc_FFT}
\end{figure*}

\begin{figure*}[ht]
       \includegraphics[width=.45\linewidth]{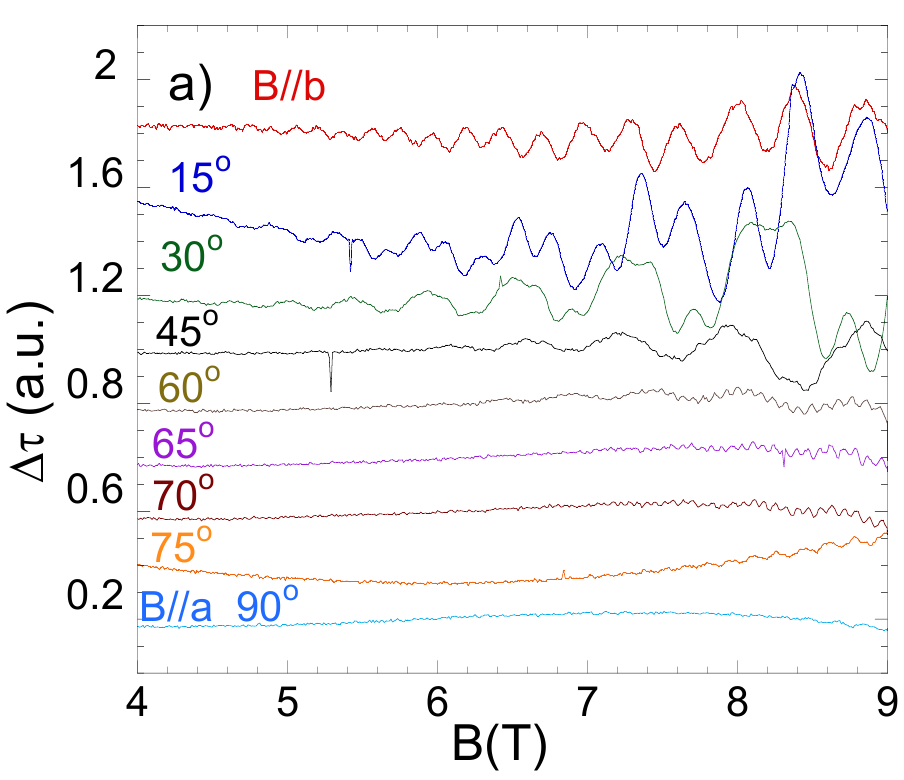}%
\hfill
       \includegraphics[width=.45\linewidth]{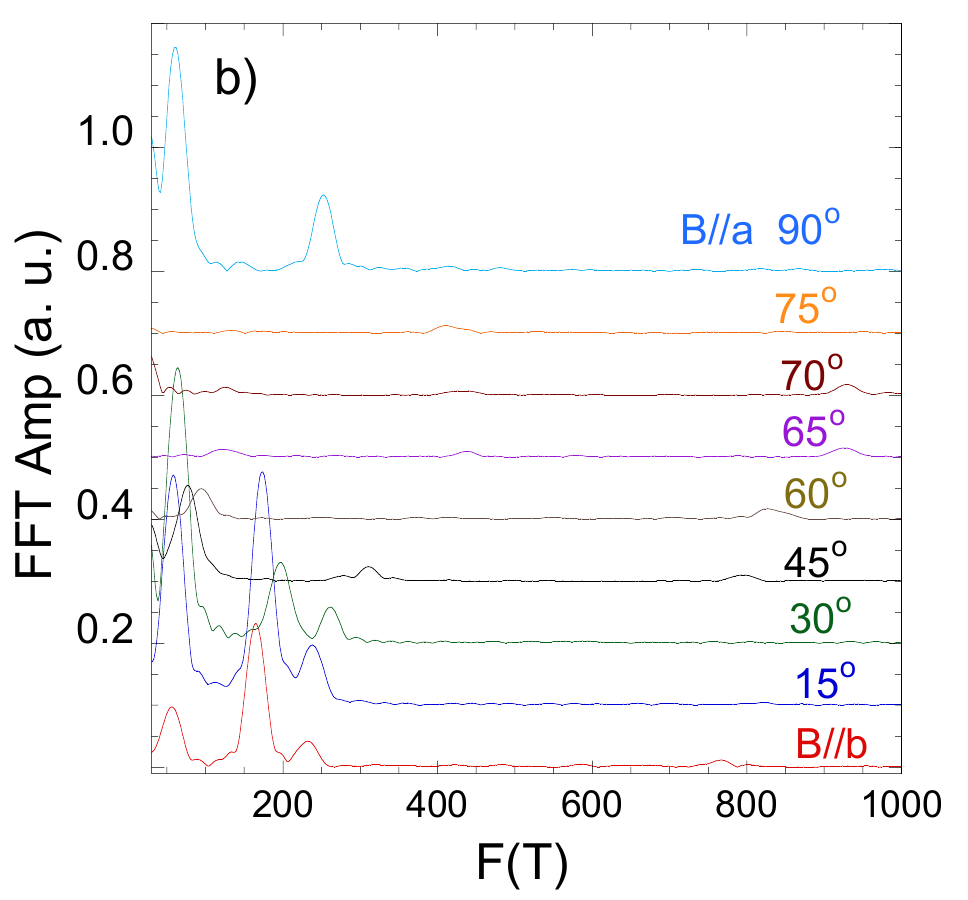}%
       \caption{a) Background subtracted magnetic torque data for \textbf{B}$\parallel$\emph{a}-\emph{b} plane at \SI{1.8}{\kelvin} and in the magnetic field range from \SI{4}{\tesla} up to \SI{9}{\tesla}. b) Corresponding FFT spectra.}
  \label{SI_backgroundsubtr+BIIab_analysis}
\end{figure*}

\begin{figure*}[ht]
       \includegraphics[width=.45\linewidth]{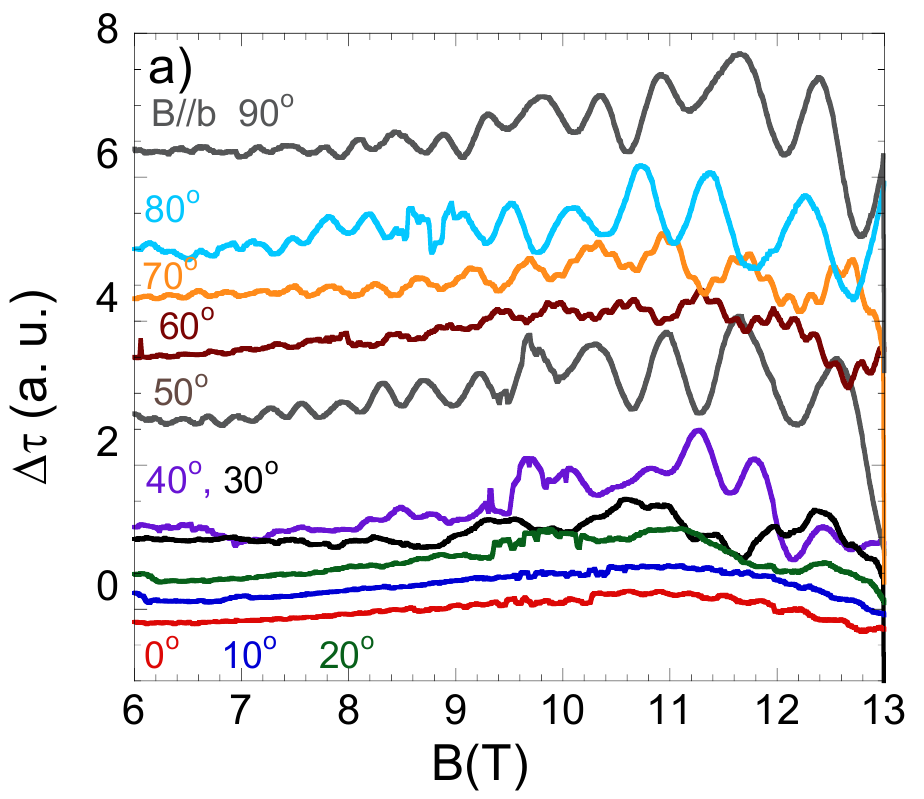}%
\hfill
       \includegraphics[width=.45\linewidth]{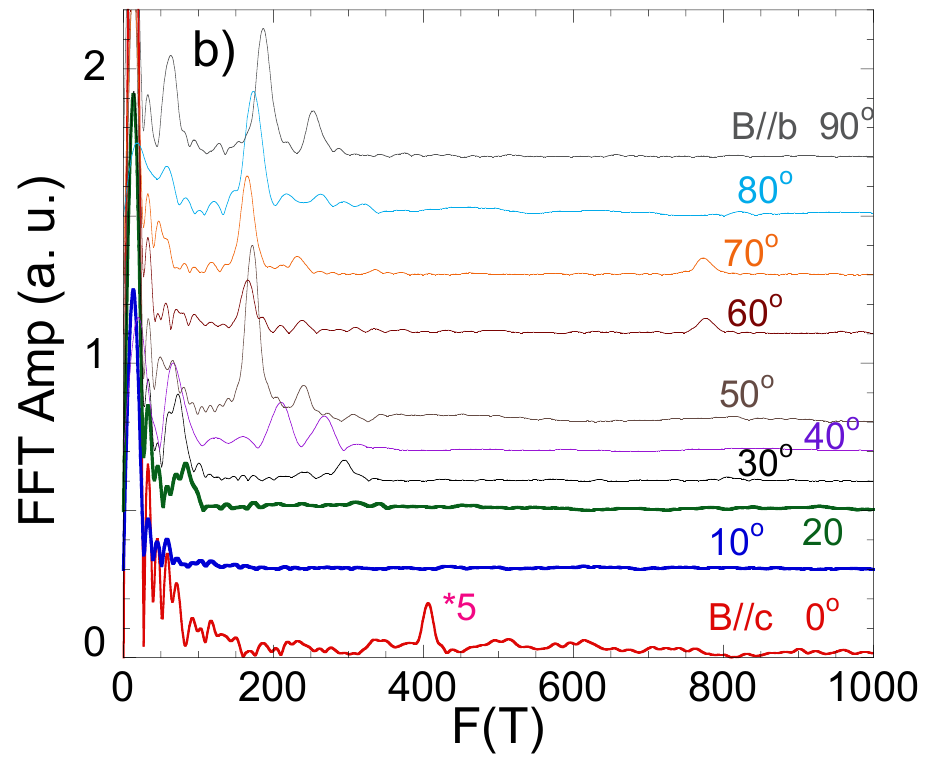}%
       \caption{a) Background subtracted magnetic torque data for \textbf{B}$\parallel$\emph{b}-\emph{c} plane at \SI{1.8}{\kelvin} and in the magnetic field range from \SI{6}{\tesla} up to \SI{13}{\tesla}. b) Corresponding FFT spectra.}
  \label{SI_backgroundsubtr+BIIbc_analysis}
\end{figure*}

\pagebreak
\section{Raw Data for Transport and Thermopower}

\begin{figure*}[ht]
        \includegraphics[width=\linewidth]{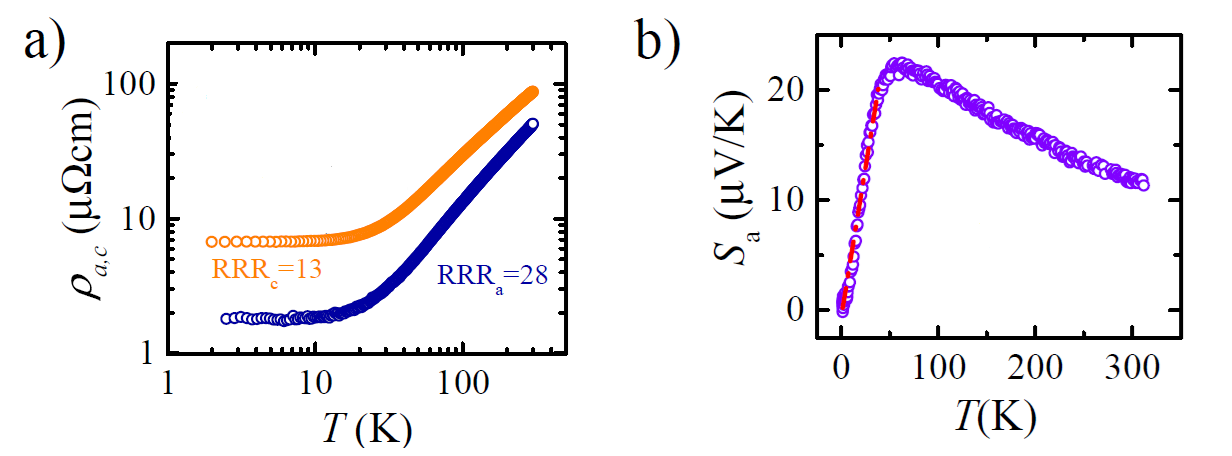}
    \caption{a) Temperature dependence of the resistivity with RRR$_a$ = 28 and RRR$_c$ = 13 in the temperature range from \SI{300}{\kelvin} down to \SI{2}{\kelvin}. b) Temperature dependence of the Seebeck coefficient along the \emph{a}-axis in the temperature range from \SI{300}{\kelvin} down to \SI{2}{\kelvin}. }
    \label{SI_res-Seeb_IIa_raw+backgroundsubtr}
\end{figure*}

\pagebreak
\section{Raw Data for Magnetoresistance}\label{raw_data_magnetoresistance}

\begin{figure*}[ht]
        \includegraphics[width=.45\linewidth]{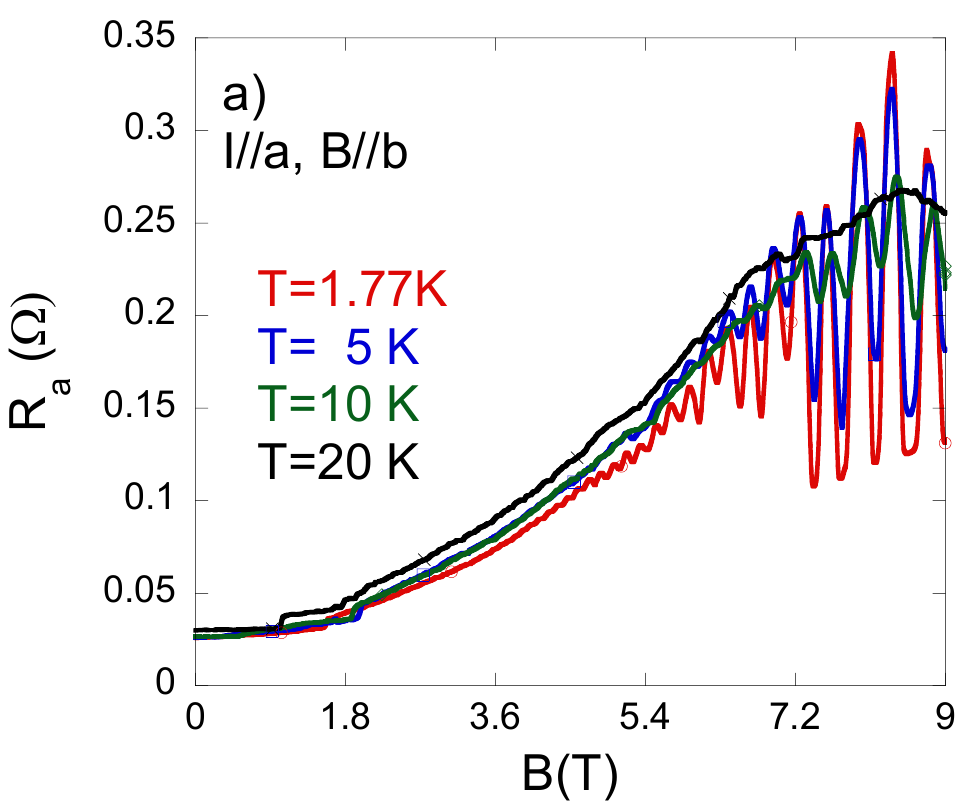}%
	\hfill
	\includegraphics[width=.45\linewidth]{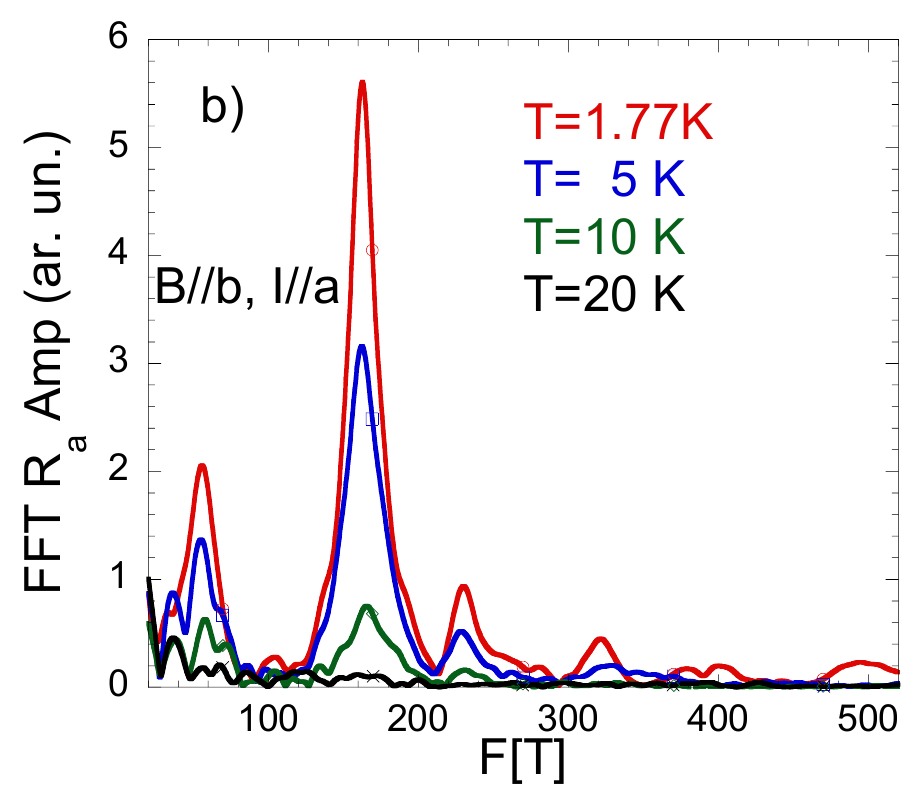}%
	\caption{a) Magnetic field sweeps of the resistance at various temperatures between \SI{1.77}{\kelvin} and \SI{20}{\kelvin}. b) Corresponding FFT spectra generated from the raw data between \SI{6}{\tesla} and \SI{9}{\tesla}.}
    \label{SI_res_BIIb_raw+backgroundsubtr}
\end{figure*}

\begin{figure*}[ht]
     \includegraphics[width=.45\linewidth]{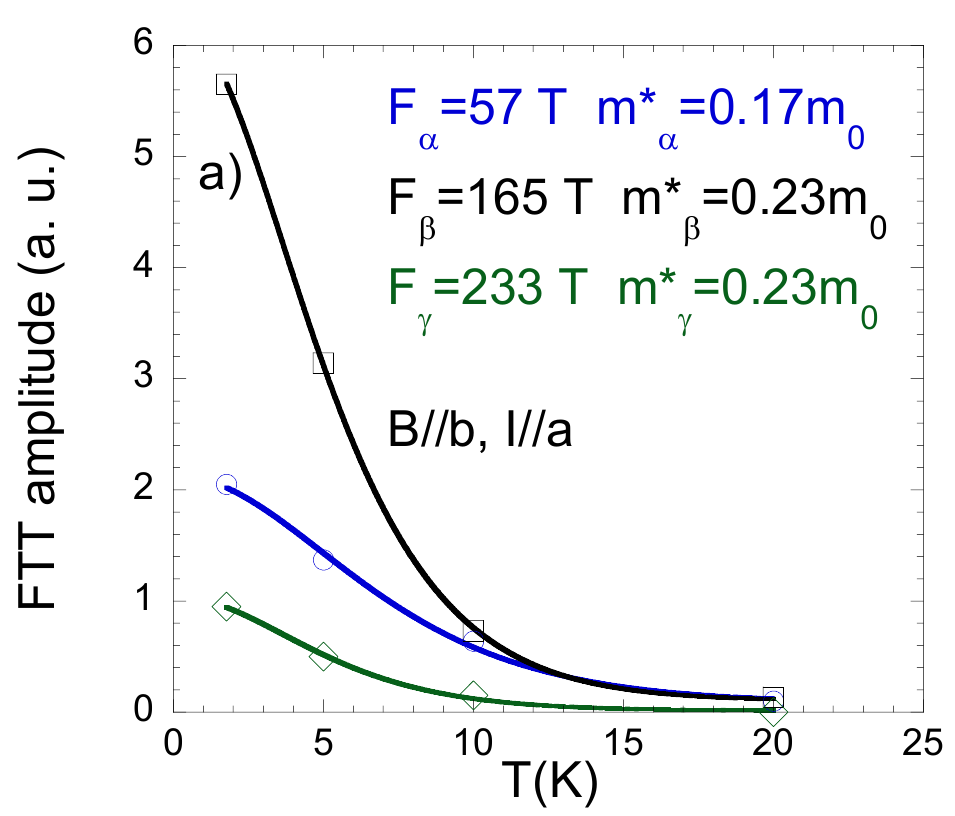}%
\hfill
        \includegraphics[width=.45\linewidth]{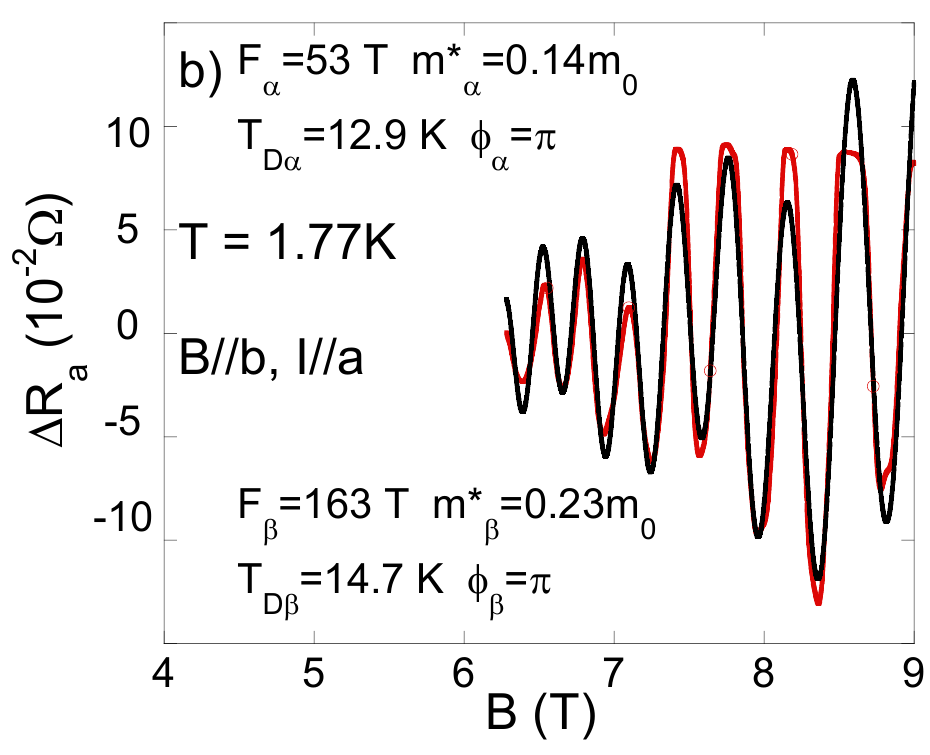}%
	\caption{a) The fits of the temperature damping factor $R_T$ in the LK formula to the FFT amplitudes.  b) The oscillatory component of the background-subtracted resistance and corresponding fit at T = \SI{1.77}{\kelvin}.}
    \label{SI_res_BIIb_FFT+effectivemass}
\end{figure*}

\end{document}